\documentclass[singlecolumn,notitlepage,prf,floatfix,amsmath,amssymb]{revtex4-2}

\usepackage{graphicx}   
\usepackage{dcolumn}    
\usepackage{bm}         
\usepackage{tikz}
\usepackage{amssymb}
\usepackage{xcolor}
\usepackage{booktabs}
\usepackage{makecell}
\usepackage{hyperref}

\begin{document}

\preprint{APS/123-QED}

\title{Laminar and Turbulent Flow in Wavy Pipes under Strong Wall Modulations}

\author{I. El Mellas}
\email{ismail.elmellas@csic.es}
\author{J.~J. Hidalgo}
\author{M. Dentz}

\affiliation{Spanish National Research Council (IDAEA-CSIC), Barcelona, Spain}

\date{\today}

\begin{abstract}
We study laminar, transitional and turbulent flow in wavy pipes
using direct numerical simulations (DNS) for bulk Reynolds numbers between
$1$ and $5300$. Flow behaviors are analysed in terms of the friction factor $f$
and mean velocity statistics for strong sinusoidal wall fluctuations
in the axial direction. Depending on the wall amplitude $k$, flow
reversal may appear at bulk Reynolds numbers as small as $25$, inducing
local recirculation zones significantly increasing friction in the
laminar regime. These effects are not captured by classical models
based on bulk geometric parameters, but require the definition of an
effective hydraulic radius $R_e$ as a hydrodynamic
concept. Furthermore, wall modulations appear to trigger supercritical
transitions to turbulence in a Reynolds range
between 500 and 1000, well below the classical
threshold for smooth pipes. The DNS data suggest an upper bound for the
laminar-turbulent transition with a critical Reynolds number that scales as $Re_c \sim
k^{-1/\gamma}$ with $5/4 \leq \gamma \leq 3/2$, consistent with
finite-amplitude transition scenarios. In the turbulent regime, flow
is found to be fully rough, dominated by inertial separation and wall-induced
disturbances independent of $Re$. Using $R_e$ as the characteristic
length scale, the wall amplitude provides a robust estimator for the
equivalent sandgrain roughness $k_s$. The
impact of strong wall fluctuations on laminar and turbulent friction
laws, as quantified by $R_e$ and $k_s$, and the
amplitude dependence of the critical Reynolds number, emphasize the
limitations of the Moody diagram for the flow quantification in
conduits with strong wall fluctuations across all flow regimes.
\end{abstract}

\maketitle

\section{\label{sec:headings}Introduction}

Most real conduits are neither straight nor smooth. Corrugated pipes,
stented arteries, ventilation ducts and fractured rocks all show
variations in cross-section, curvature and surface texture that alter
flow resistance and mixing. Karst caves represent an extreme natural
example: dissolution of limestone produces channels with irregular
shapes and rough walls \cite{curl1966caves, dreybrodt2012processes,
claudin2017dissolution}. These geometric variations control head loss,
solute transport and residence time, yet hydrological models still
rely on empirical formulas derived for smooth pipes, such as the
Hagen-Poiseuille law or the Darcy-Weisbach equation
\cite{cohen2004fluid}. When roughness is considered, it is typically
introduced through a relative roughness correction using the Moody
diagram \cite{moody}, which assumes a uniform distribution of
small-scale texture superimposed on an otherwise circular
cross-section. This approach becomes increasingly unreliable when the
roughness is large enough to notably alter the cross-sectional area itself, as
is often the case in natural conduits. In such conditions, roughness
is no longer just a wall property but a defining part of the overall
geometry, and the underlying flow physics cannot be captured by
standard friction factor correlations.

To isolate the influence of wall geometry, many studies use simplified
rough-walled pipes.  Such configurations allow systematic changes in
roughness height, wavelength or curvature, providing a bridge between
smooth, idealized models and the multiscale geometry of natural
conduits. Many studies have aimed to define an equivalent sand-grain roughness
height ($k_s$) to represent the effect of different surface roughness
geometries. The equivalent sand-grain roughness height, $k_s$,
introduced by \citet{nikuradse1933} and \citet{schlichting1936},
remains the principal reference for quantifying wall friction. The
non-dimensional roughness height is typically expressed in viscous
units as $k_s^+ = k_s u_\tau/\nu$, where $\nu / u_\tau$ is
the viscous length scale, $\nu$ is the kinematic viscosity and
$u_{\tau} = \sqrt{\tau_w/\rho}$ is the friction velocity.

Following \citet{Flack2014}, flow is hydrodynamically smooth when
$k_s^+ \lesssim 5$, transitionally rough for $5 \lesssim k_s^+
\lesssim 70$ and fully rough when $k_s^+ \gtrsim 70$.  Roughness
increases wall friction compared with smooth walls, but the mechanisms
differ across these regimes: viscous damping dominates in the smooth
regime, geometry controls drag in the transitional regime, and
pressure drag governs friction in the fully rough regime, making it
nearly independent of Reynolds number.

Compared to smooth surfaces, wall roughness enhances friction, which
results in a reduced inner-scaled mean velocity profile. This
reduction, referred to as the roughness function proposed by
\citet{clauser1954} and \citet{hama1954}, is typically expressed as:
\begin{equation}
    \Delta U^+ = \frac{1}{\kappa} \log k_s^+ + A - B(k_s^+),
    \label{du}
\end{equation}
where $\kappa \approx 0.4$ is the von K\'{a}rm\'{a}n constant, $A \approx
5$ is the smooth-wall additive constant, and $B$ depends on the
roughness geometry and $k_s^+$ \cite{jimenez2004}. As $k_s^+$
increases, the flow moves from the smooth to the transitional
regime. When $\Delta U^+ > 7$, the flow is usually considered fully
rough, and $B$ becomes independent of Reynolds number.
In wall-bounded turbulent flows, the velocity profile can be
separated into two regions: the inner layer, which is directly
influenced by the wall and viscous effects; and the outer layer, where
the dominant motions are larger in scale and relatively unaffected by
the fine-scale details of the wall. The inner layer includes the
viscous sublayer, the buffer region, and the lower logarithmic layer,
and its structure is strongly modified by roughness elements,
particularly when their height in wall units, $k_s^+$, becomes
significant.

In contrast, the outer layer is assumed to become increasingly
independent of wall characteristics as the Reynolds number
increases. This is the basis of Townsend's outer-layer similarity
hypothesis \cite{Townsend1976}, which states that when there is
sufficient scale separation (i.e., $k_s/\delta \ll 1$, where $\delta$
is the outer length scale of the flow, such as the pipe radius),
roughness effects remain confined to the near-wall region. Under these
conditions, the outer-layer turbulence retains a universal structure,
and the effect of roughness manifests primarily as a downward shift in
the mean velocity profile, quantified by an additive constant $\Delta
U^+$, in the logarithmic law. However, this assumption may break down
when the roughness becomes comparable to the boundary layer thickness
or induces large-scale secondary flows that extend into the outer
region \cite{jimenez2004}. Townsend's hypothesis has been shown to
approximately hold across a variety of canonical roughness types, such
as sand-grain roughness, cylindrical elements, and spherical particles
\cite{Flack2005, Chan2015, busse2015, busse2017}. These studies, both
experimental and numerical, represent roughness as
small-scale geometric perturbations superimposed on otherwise smooth
and straight channels or pipes, using idealized elements like
hemispheres, ribs, or random sand-grain topographies
\cite{jimenez2004, busse2015, thakkar2016}.

More recent investigations have
extended these works to irregular and stochastic roughness in both plane
channels and pipes, exploring how roughness amplitude, spacing, and
spatial arrangement modulate near-wall turbulence and affect the mean
flow \cite{thakkar2018, demaio}.
Parallel efforts have aimed to establish correlations between
the equivalent sand-grain roughness height $k_s$ and the underlying
geometric characteristics of the roughness elements 
\cite{napoli2008, Chan2015, saha_2015, Forooghi2017, Flack2020, nair2025rough}.
However, rough-wall models typically assume statistical homogeneity in
the axial direction, with wall deformations treated as local surface
features rather than large-scale geometric variations. In many natural
systems, such as karst conduits, the geometry changes more
fundamentally along the flow path, affecting the cross-section,
curvature and flow alignment in ways that local roughness models
cannot capture \cite{Jeannin2001}.

To explore these effects in a simplified and
controlled setting, a widely used configuration is the wavy pipe.
It offers an intermediate step between rough pipes and fully irregular
natural conduits that is able to reveal the salient features of the flow
dynamics in natural conduits. For example, while real karst conduits
are more irregular and non-periodic,  the wavy-wall geometry serves
as a canonical model capturing generic mechanisms associated with wall
undulations, rather than a geometrically faithful representation of
karst systems.

\citet{hsu_1971} and \citet{nishimura1984}
examined flow through channels with sinusoidal walls, showing that wall waviness promotes
vortex formation and increases pressure losses. The former focused on
turbulent flow in wavy pipes, whereas the latter investigated steady,
quasi-two-dimensional laminar flow. In both cases, the recirculating
vortices that develop in the diverging sections raise wall shear and
pressure drop. \citet{pozrikidis1987} studied creeping Stokes flow in
two-dimensional periodic channels and demonstrated that flow reversal
and recirculation can arise even at vanishingly small Reynolds
numbers, with a strong dependence on the driving mechanism and channel
shape. Complementing these works, \citet{nishimura1995} carried out
experiments on oscillatory flow and mass transfer in both asymmetric
and symmetric sinusoidal channels, showing that the onset of unsteady
oscillations markedly enhances mass-transfer
rates. \citet{Sisavath_2001} and \citet{bernabe2000hydraulic}
studied the hydraulic conductivity of creeping flow in wavy pipes.

\citet{wang2008} used the lattice Boltzmann method to examine viscous flow in wavy
pipes, for which they observed secondary flows and enhanced pressure losses driven
by curvature. \citet{cotrell2008instability} and \citet{loh_2011}
performed linear stability analyses of flow in circular pipes
with sinusoidal axial corrugations. The analysis of
\citet{tao2009critical} shows that even for fine corrugations, $3-4$
orders of magnitude smaller than the pipe diameter, the flow becomes
linearly unstable at moderate Reynolds numbers. These works show that wall
corrugations can lead to linear instability at finite Reynolds
numbers, qualitively different from smooth pipes, for which flow is
linearly stable.   
This promotes transition from stable to unstable flow at
smaller Reynolds numbers than for smooth pipes. 
The experiments of \citet{deiber_1979} and \citet{nishimura_2003} of turbulent flow in
wavy pipes have observed the onset of turbulence at Reynolds number way below
the value of about $2000$ for smooth pipes. These findings
suggest wall waviness can significantly destabilize laminar pipe
flow. The work of \citet{saha_2015} analysed pipe flows over sinusoidal transverse
corrugations for relatively moderate corrugation heights and proposed scaling laws to quantify friction and flow
structure across laminar to transitional regimes.

While these studies have advanced the understanding of flow
in wavy or corrugated geometries, important gaps remain. Direct
numerical simulations (DNS) exploring strong
geometric variations, especially at high amplitude-to-diameter ratios,
are scarce \cite{saha_2015}. Moreover, few works have systematically
examined how such geometries influence the transition from laminar to turbulent flow
\cite{cotrell2008instability, loh_2011}.
The definition of an effective hydraulic diameter in non-uniform geometries remains
ambiguous, complicating the use of conventional scaling
laws. Similarly, attempts to quantify an equivalent sand-grain
roughness height for globally deformed yet smooth walls are limited,
leaving friction factor comparisons with canonical rough-wall data,
such as Nikuradse's, on uncertain ground.

We address these open questions by analysing flow in circular pipes
with axisymmetric wavy walls using DNS. The pipe radius varies
periodically along the streamwise direction. The wall geometry is
defined analytically, enabling
controlled variation of roughness amplitude and wavelength while
preserving a smooth, circular cross-section. The analysis spans
laminar, transitional, and turbulent regimes to examine how
wave-induced roughness influences flow development, pressure drop, and
the onset of turbulence. Particular focus is given to the
characterization of wall friction and the definition of an effective
sand-grain roughness height. This enables comparison not only with
classical data such as Nikuradse's fully rough pipe measurements, but
also with more recent studies involving controlled roughness
geometries \cite{nishimura1984, saha_2015, demaio}, who examined
higher roughness levels in random and wavy rough pipes. By varying
Reynolds number and geometric parameters systematically, the study
aims to clarify the role of large-scale, structured roughness in both
momentum transport and flow regime transitions.

\section{\label{sec:methodology}Methodology}

In the following, we define the pipe geometry, state the flow equation
and basic relations of pipe flow, and provide
details of the direct numerical simulations (DNS).

\begin{figure}[t]
\includegraphics[width=.9\columnwidth]{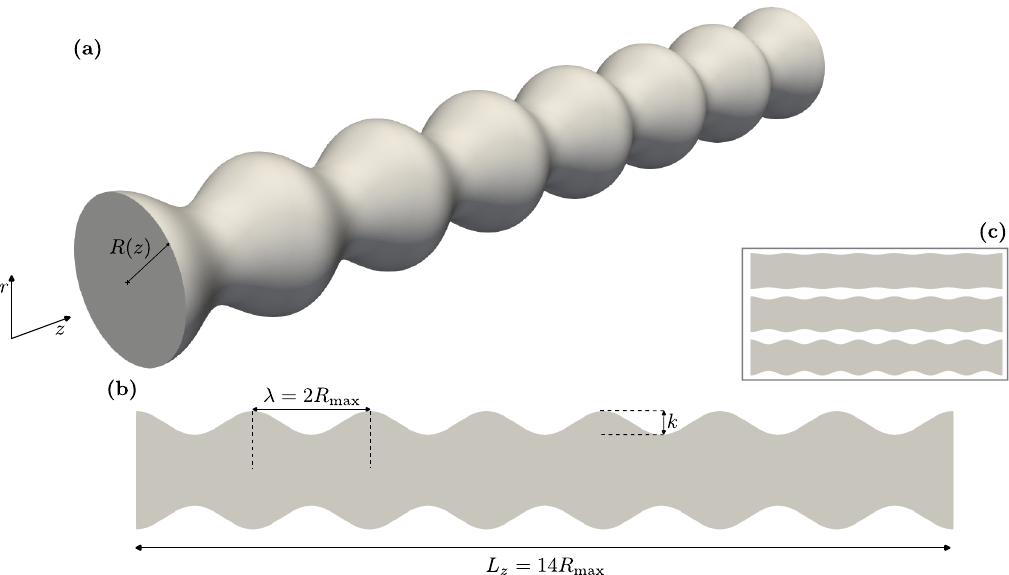}
\caption{Schematic of the simulation geometry.
    (a) Three-dimensional view of the wavy pipe with the strongest wall modulation ($\overline{R}/k=2$), where the wall radius
    varies periodically along the axial direction according to Eq.~\eqref{r}.
    (b) Longitudinal cross-section of the wavy pipe in (a) showing the full pipe length
$L_z = 14R_{\text{max}} = 7\lambda$, with one wavelength $\lambda$ of the wall modulation indicated. $k$ represents the peak-to-peak amplitude of the wall waviness. (c) Longitudinal cross-sections of the remaining wavy geometries, $\overline{R}/k=10, 5, 3.3$ (top to bottom). 
The flow is oriented along the $z$-axis. Periodic boundary
conditions are imposed at the inlet and outlet, while no-slip conditions are applied at the pipe wall. .}
\label{smooth_geo}
\end{figure}

\subsection{\label{sec:geometry}Pipe geometry}

We consider the wavy pipe geometry illustrated schematically in
Fig.~\ref{smooth_geo}, which also defines the notation used in the
following. The wall radius $R(z)$ measured from the straight centerline is given
by the analytical expression
\begin{align}
    R(z) = R_{\max} -\frac{k}{2}\left[1-  \sin\left(\frac{2\pi
    z}{\lambda}\right)\right].
    \label{r}
\end{align}
Here $R_{\max}$ is the maximum pipe radius, $k < R_{\max}$ is the peak-to-peak
amplitude of the wall waviness, i.e., $R_{\min} = R_{\max} - k$, and
$\lambda$ the streamwise wavelength of the sinusoidal
modulation. We fix here $R_{\max}$ and $\lambda = 2 R_{\max}$ and vary
in the following only $k$. The baseline smooth
pipe is characterized by the radius $R(z) = R_{\max}$. The average pipe
radius is given by $\overline R = R_{\max} - k/2$. The cases considered in
this work span the five values $k = 0,\ 0.0477\lambda,\ 0.091\lambda,\ 0.1304 \lambda$, and
$0.2 \lambda$, which correspond to $\overline R/k = 0, 10, 5, 10/3, 2$, see
also Table~\ref{table:1}. While $R_{\max}/k$ is an equally valid 
geometric descriptor, we adopt $\overline{R}/k$ as it varies 
continuously from zero in the smooth-wall limit, providing a natural 
parametrization of the wall modulation amplitude relative to the mean 
pipe radius. The variable pipe diameter is denoted by
$D(z) = 2 R(z)$. The position vector is denoted in Cartesian
coordinates by $\mathbf x = (x,y,z)^\top$, where $\top$ denotes the
transpose.

Throughout this work, the terms waviness and
roughness are used interchangeably, although the imposed geometry
is smooth and periodic.
The geometrical roughness height here is defined as the peak-to-peak wall
amplitude, $k$. Other definitions for the wall roughness are discussed
in Sec.~\ref{sec::ks}. As emphasized by \citet{jimenez2004},
roughness or equivalent sand-grain roughness is a hydrodynamic concept
that reflects the impact of wall fluctuations on wall shear stress and
flow resistance. A challenge here is to identify and relate the geometric wall
characteristics to the flow resistance in the fully rough regime
\cite{chung2021}, which is discussed in Sec.~\ref{sec::ks}. In the
following, we summarize a series of geometric surface statistics that
have been proposed to characterize geometric roughness with the aim of
correlating them to the hydrodynamic function of roughness.

\citet{schlichting1936} defines the solidity $\sigma$ to characterize the
surface geometry as the ``total projected frontal area of a roughness
element per unit wall-parallel projected area''~\cite{jimenez2004}. For
the wavy pipes under consideration here it is $\sigma = k/\lambda$.
\citet{napoli2008} propose the effective slope $ES$,
\begin{equation}
    ES=\frac{1}{\lambda} \int_0^{\lambda} \left|\frac{d
    R(z)}{dz} \right| dz,
\end{equation}
as a measure that is equivalent to solidity, but simpler to use for
random surfaces. In fact, the effective slope is two times the
solidity $ES = 2 \sigma$.  That is, here $ES = 2 k/\lambda$. Other
authors \cite{Chan2015, Forooghi2017, Flack2020} consider surface
statistics of the roughness height fluctuation \cite{musker1980},
which here is simply
\begin{align}
k'(z) = R(z) - \overline R = \frac{k}{2} \sin\left(\frac{2\pi
z}{\lambda}\right).
\end{align}
Those authors use the skewness $S_k$ of the height fluctuation, which here
is exactly zero, the root-mean square roughness $S_q$, which here is
$S_q = k/2\sqrt{2}$, or the average absolute height $S_a$, which here
is $S_a = k/\pi$.

\subsection{\label{sec:flow}Flow}

The flow is governed by the incompressible Navier-Stokes equations
\begin{equation}
\nabla \cdot \boldsymbol{u}=0
\label{mass}
\end{equation}
\begin{equation}
 \frac{\partial  \boldsymbol{u}}{\partial
t}+\boldsymbol{u}\cdot\nabla\boldsymbol{u} =-\frac{1}{\rho}\nabla
p+\nu\nabla^{2}\boldsymbol{u}+\boldsymbol{F}(t),
 \label{momentum}
\end{equation}
where $\boldsymbol{u} = (u_x, u_y, u_z)^\top$ denotes the fluid velocity
vector and velocity components in $x, y$ and $z$-directions of the
coordinate system, $t$ the time
and $p$ the pressure. A spatially uniform forcing term
$\boldsymbol{F}(t) = [0,0,F_z(t)]^\top$, is applied in the streamwise
direction to represent the mean pressure gradient while allowing axially periodic boundary conditions.
In the present simulations $F_z(t)$ is adjusted dynamically so as to
maintain a constant bulk velocity. No-slip boundary conditions are
applied at the walls of the pipe and
periodic boundary conditions at the inlet and outlet. The bulk
velocity $u_b$ is imposed, which is defined as the spatial average of the
time-averaged streamwize velocity $u$
\begin{align}
 u_b = \frac{1}{V}\int_\Omega u \,dV, \qquad u = \frac{1}{T} \int_0^T
u_z \,dt,
\label{bulk_vel}
\end{align}
where $\Omega$ denotes the flow domain, $V$ its volume, and $T$ the
averaging interval discussed in Sec.~\ref{sec:time} below.

The Reynolds number is defined in terms of a characteristic radius
$R_c$, which is discussed in Sec.~\ref{sec:lf}, and the bulk velocity $u_b$ as
\begin{equation}
\label{eq:Reb}
    Re=\frac{2R_{c} u_b}{\nu}.
\end{equation}
The pressure gradient $G=-dP/dz$ is a result of
the solution of the Navier-Stokes equation. The wall shear stress
$\tau_w$ is defined in terms of $G$, the friction velocity $u_\tau$ in
terms of $\tau_w$ as,
\begin{align}
\label{eq:tauw}
\tau_w=\frac{R_{c}G}{2}, \qquad u_\tau = \sqrt{\frac{\tau_w}{\rho}}.
\end{align}
We use a constant characteristic radius to define the wall shear stress.
Note, however, that the wall shear stress varies in streamwise
direction along the wavy pipe. The friction velocity $u_\tau$ is used
to define the friction Reynolds number as
\begin{align}
\label{eq:Retau}
Re_{\tau}= \frac{R_c u_\tau}{\nu}.
\end{align}
The roughness Reynolds number $k^+$ is defined as
\begin{align}
k^+ = \frac{k u_\tau}{\nu}.
\label{eq:kpiu}
\end{align}
It compares the geometric roughness height $k$ and the viscous length
scale $\delta = \nu/u_{\tau}$.  For $k^+ \gg 1$, the wall roughness
height is penetrating through the viscous wall layer. In this fully
rough regime, wall shear stress becomes independent of viscosity. Wall
friction is dominated by Reynolds stresses and turbulent pressure drag
across the boundary layer. In the following, quantities that are rescaled in terms of
wall units (velocities by $u_\tau$, length by $\delta$) are denoted by
a superscript $^+$. Furthermore, we define the bulk Reynolds number in
terms of the maximum radius as $Re_b = 2 R_{\max} u_b/\nu$. Recall
that $R_{\max} = \lambda/2$ has the same value for all configurations
under consideration.

\subsubsection{\label{sec:darcy}Darcy friction factor}

The global flow resistance due to wall friction is quantified by the
Darcy friction factor. It is defined according
to the Darcy-Weisbach formulation as \cite{white2011}
\begin{equation}
  f=\frac{4 R_{c}}{\rho u_b^{2}}G.
 \label{f_eq}
\end{equation}
In terms of the wall shear stress $\tau_w$ it reads as
\begin{equation}
    f=8\tau_w/(\rho u_b^2).
\end{equation}
The friction factor is used throughout this study as a global
indicator of flow resistance. While in smooth-wall flows $f$ depends
primarily on Reynolds number, geometric modifications such as wall
waviness introduce additional drag via the creation of recirculation
bubbles, enhanced turbulence production and form-induced pressure
losses. The evolution of the friction factor
with increasing Reynolds number and wall modulation will be used to
quantify the hydrodynamic impact of waviness and to establish
comparisons with classical rough-wall behavior. The friction factor
defined in terms of the bulk Reynolds number $Re_b$ is denoted by
$f_b$.

\subsubsection{\label{sec:velstats}Velocity statistics}

In order to assess the impact of roughness on the turbulence intensity
and the locations of turbulence production, we focus on profiles of the
mean streamwise velocities denoted in the following by $u_z$. The variances of the time-averaged
streamwise, radial and angular velocity components, denoted by $u_r$
and $u_\theta$ are reported in Appendix \ref{app}. We denote the axial,
radial and angular coordinates by $z, r$ and $\theta$, respectively.

Due to the variable wall geometry, we depart from the standard
practice of averaging over the entire axial domain.
The analysis of individual regions allows for the identification of
local flow behaviors, such as recirculation and flow detachment,
which would otherwise be obscured by a global average.
Thus, we extract cross-sectional
slices at selected streamwise locations that correspond to distinct
geometric features of the wavy wall profile, as defined
in~Eq.~\eqref{r}. Specifically, we select contraction regions where the
wall radius reaches its minimum, $R(z_i) = R_{\min}$, at $z_i = (2 i -
1)\lambda/2$, expansion regions where the radius is maximal, $R(z_i) = R_{\max}$, at
$z_i = i \lambda$, and intermediate regions where the radius takes the average value $R(z_i) = \overline R$, at $z_i = (2 i - 1)\lambda/4$.
Thus, the average for each region along the pipe is defined by
\begin{align}
\langle \phi \rangle_k(r) = \frac{1}{N_k} \sum_{i = 1}^{N_k}
\frac{1}{2\pi}\int_0^{2\pi} d \theta\,
\phi(r,z_i,\theta),
\end{align}
where $k = \min, \max, \text{mid}$ indicates the location of interest at
$R(z) = R_{\min}, R_{\max}, R_{\text{mid}} = \overline R$; $N_k$ is
the respective number of cross-sections, and $\phi$ the target variable, that is,
the velocity components and their squares. To enable consistent
comparison between different axial regions despite local variations in
pipe radius, the radial coordinate is rescaled relative to the maximum
radius $R_{\max}$ such that the rescaled radial coordinate $r'$ is
defined as
\begin{align}
r' = \frac{r}{R(z)} R_{\max}.
\end{align}
This transformation maps all radial profiles onto a common reference
frame. Furthermore, we display the results in terms of the distance
from the wall, which is given by $y = R_{\max} - r'$, and in terms of
the distance in wall units, which is $y^+ = y u_\tau/\nu$.

The global average is obtained by averaging the profiles at the same
rescaled wall distance as
\begin{align}
\label{eq:average}
\langle \phi \rangle (y) = \frac{1}{2 \pi L_z} \int_0^{L_z} dz
\int_0^{2\pi} d \theta \,\phi[(R_{\max} - y)R(z)/R_{\max}].
\end{align}
That is, we average velocities at the same relative distance to the
wall. Specifically for strong wall fluctuation, this accounts for the
fact that the position of the boundary layer may vary along the wavy
wall contour.
Comparing these global averages with the local profiles at
characteristic cross-sections allows quantification of dispersive
stresses, which arise from spatial heterogeneity in the time-averaged
flow \cite{Nikora_2007b}.
In the following, we employ for the average streamwise velocity
component the notation $U(y) \equiv \langle u_z \rangle$, and in wall
units $U^+(y^+) \equiv U(y^+ \nu/u_\tau)/u_\tau$.

\subsubsection{\label{sec:loglaw}Logarithmic laws and roughness function}

Pipes are considered smooth if the roughness height is smaller than
the viscous sublayer, $k^+ \ll 1$. For large $Re \gg 1$ an inertial
range develops for $y \gg \nu/u_\tau$ in which the only relevant
length scale is the distance to the wall. The average velocity profile
for $y^+ \gg 1$, is then given by the logarithmic law \cite{cohen2004fluid}
\begin{align}
\label{eq:Usmooth}
U^+ = \kappa^{-1} \ln(y^+) + A,
\end{align}
where $\kappa = 0.4$ is the von K\'{a}rm\'{a}n constant, the constant $A$ is
typically found to be $A \approx 5$. For rough surfaces ($k^+ \gg 1$) flow behaves similarly
to smooth surfaces at distances $y \gg k$ large compared to a
characteristic roughness length $k_s$ \cite{Townsend1976}, and follows \cite{chung2021}
\begin{align}
\label{eq:Urough}
U^+ = \kappa^{-1} \ln(y/k_s) + B(k_s^+),
\end{align}
where $B$ depends on the wall geometry and roughness Reynolds
number. For $k_s^+ \ll 1$ it tends toward a constant asymptotic value
$B^\infty \equiv B(\infty)$. The characteristic length $k_s$ depends on the wall geometry,
but it is a hydrodynamic concept \cite{jimenez2004, chung2021}.
For rough surfaces, the mean velocity decreases with respect to the
one for smooth surfaces at the same or similar $Re_\tau$. This
velocity deficit is quantified by the roughness function $\Delta U^+$. We write the mean velocity for the wavy pipe as \cite{chung2021}
\begin{align}
\label{eq:deltaU}
U^+ = \kappa^{-1} \ln(y^+) + A - \Delta U^+, \qquad \Delta U^+ =
\kappa^{-1} \ln(k_s^+) + A - B(k_s^+).
\end{align}
We use the roughness function $\Delta U^+$ in order to characterize and
measure the impact of roughness on the mean flow behavior. As
outlined above, for rough pipes, the equivalent sand-grain roughness $k_s^+$ has been used as a
hydrodynamic concept to quantify the impact of wall roughness on
flow. It can be defined by inverting the Colebrook relation for the
velocity deficit \cite{jimenez2004, thakkar2018, demaio}
\begin{equation}
    \Delta U^+ = \kappa^{-1} \ln(1 + 0.3 k_s^+).
    \label{colebrook}
\end{equation}
Sand-grain roughness needs to be related to the actual surface geometry to be used as
a meaningful roughness descriptor. We will discuss this question in
Sec.~\ref{sec::ks}. Note that using the Colebrook relation
Eq.~\eqref{colebrook} fixes $B^\infty$ to the value $B^\infty = A -
\kappa^{-1} \ln(0.3)$.

As discussed by \cite{raupach1991} and \cite{chung2021} care needs to
be taken regarding the origin of $y$ in Eq.~\eqref{eq:Urough} because
the wall modulation moves the entire flow toward the center of the
conduit, which is not negligible for the wall amplitudes under
consideration. That is, the logarithmic law
Eq.~\eqref{eq:Urough} is expected to hold for the velocity as a function
of $Y = y - \ell$, where $\ell$ is denoted the wall-offset or position
of the virtual origin. The wall-offset used in this work
is discussed in Appendix~\ref{app:offset}.

\begin{table*}[b]
  \caption{Flow parameters for flow in wavy pipes. The table is divided
in two sets of bulk Reynolds number $Re_b=2R_{\max}u_b/\nu$,
respectively 2000 and 5300. $\overline{R}/k$ is the inverse relative
roughness, comparing the average radius $\overline{R}$ and the wall roughness $k$. Minimum
$\Delta r^{+}$, $\Delta \theta^{+}$ and $\Delta z^{+}$ are the grid spacing in the
cross-stream and streamwise direction, in wall units, for the mesh
with $145\times113\times881$ grid points in the radial, azimuthal and
axial direction respectively. The $Re_b$ is the bulk
Reynolds number, defined with maximum aperture. The
$Re=2R_hu_b/\nu$ is the Reynolds number in terms of the effective
hydraulic radius defined by Eq.~\eqref{hydraulic_r},
$Re_{\tau}=R_hu_{\tau}/\nu$ is the friction Reynolds
number. $T/\tau_t$ is the non-dimensional averaging time expressed in
eddy turnover units, where $\tau_t=R/u_{\tau}$. $f=8\tau_w/(\rho u_b^2)$ is the friction factor.}
  \label{table:1}
\begin{ruledtabular}
\begin{tabular}{lcccccccccc}
Case & Symbol & $\overline{R}/k$ & $\Delta r^{+}_{\min}$ & $\Delta \theta^{+}$ & $\Delta z^{+}$ & $Re_b$ & $Re$ & $Re_{\tau}$ & $T/\tau_t$ & $f\times10^2$ \\
\hline
Smooth pipe   & \tikz\draw[dashed,line width=2pt] (0,0) -- (0.6,0); & $-$ & 0.032 & 0.46 & 0.89 & 2000 & 2000 & 32 & 63.24 & 3.2 \\
2k-10 & \tikz\draw[red, line width=1pt] (0.1,0.1) circle [radius=0.1]; & 10 & 0.038 & 0.51 & 1.03 & 2000 & 1886 & 68.58 & 77.16 & 4.23 \\
2k-5  & \tikz\draw[blue,fill=none, line width=1pt] (0,0) rectangle (0.2,0.2); & 5 & 0.065 & 0.84 & 1.80 & 2000 & 1687 & 103.28 & 72.61 & 11.99 \\
2k-3.3 & \tikz\draw[teal,fill=none,line width=1pt] (0,0) -- (0.1,0.2) -- (0.2,0) -- cycle; & 10/3 & 0.0897 & 1.10 & 2.47 & 2000 & 1488 & 121.03 & 65.59 & 21.17 \\
2k-2  & \tikz\draw[violet,fill=none, line width=1pt] (0.1,0) -- (0.2,0.1) -- (0.1,0.2) -- (0,0.1) -- cycle; & 2 & 0.139 & 1.57 & 3.83 & 2000 & 1112 & 125.46 & 60.88 & 40.73 \\
\hline
Smooth pipe   & \tikz\draw[dashed,line width=2pt] (0,0) -- (0.6,0); & $-$ & 0.093 & 1.31 & 2.55 & 5300 & 5300 & 181.4 & 61.62 & 3.7 \\
5k-10 & \tikz\draw[red, line width=1pt] (0.1,0.1) circle [radius=0.1]; & 10 & 0.123 & 1.66 & 3.39 & 5300 & 4997 & 225.13 & 76.47 & 6.49 \\
5k-5  & \tikz\draw[blue,fill=none, line width=1pt] (0,0) rectangle (0.2,0.2); & 5 & 0.191 & 2.46 & 5.27 & 5300 & 4470 & 302.62 & 80.29 & 14.67 \\
5k-3.3 & \tikz\draw[teal,fill=none,line width=1pt] (0,0) -- (0.1,0.2) -- (0.2,0) -- cycle; & 10/3 & 0.257 & 3.17 & 7.09 & 5300 & 3943 & 345.05 & 70.57 & 24.5 \\
5k-2  & \tikz\draw[violet,fill=none, line width=1pt] (0.1,0) -- (0.2,0.1) -- (0.1,0.2) -- (0,0.1) -- cycle; & 2 & 0.387 & 4.39 & 10.68 & 5300 & 2947 & 354.52 & 56.26 & 46.31 \\
\end{tabular}
\end{ruledtabular}
\end{table*}

\subsection{\label{sec:dns}Numerical simulations}

The direct numerical flow simulations are performed using the spectral
element code Nek5000 \cite{nek}, which numerically integrates the
incompressible Navier-Stokes equations~\eqref{momentum}.
The pipe length is set to $L_z = 14 R_{\max}$. This choice follows
the recommendation by \cite{wu_2008}, who indicate that the maximum
wavelength in pipe flow is between $8$ and $16$ times the pipe
radius.

\subsubsection{\label{sec:mesh}Mesh and numerical setup}

\begin{figure}[t]
\includegraphics[width=0.5\columnwidth]{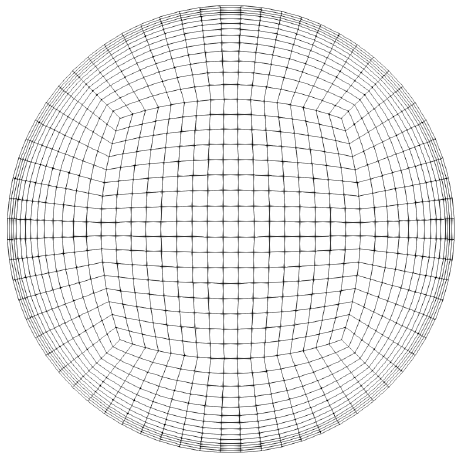}
\caption{Illustration of the hexahedral mesh. On the spectral grid the velocity is discretized using Lagrange polynomials of order
$N=9$ on Gauss-Lobatto-Legendre points, while pressure is represented
using a $P_{N-2}$ formulation.}
\label{fig:mesh}
\end{figure}

To discretize the geometry, a structured multi-block hexahedral mesh
is employed. The mesh is specifically designed to resolve the
near-wall region adequately while maintaining sufficient accuracy in the pipe
centre. The mesh is generated in Gmsh \cite{geuzaine2009gmsh} as a
structured multi-block hexahedral grid for a straight circular pipe
using transfinite curves and surfaces with recombination as
illustrated in Figure \ref{fig:mesh}. The 3D mesh
is obtained by extruding a straight circular cross-section in the
streamwise direction. The circumferential resolution is controlled by
the number of nodes along each arc, while the radial resolution is set
by two transfinite progressions: a near-wall block with strong
refinement and an interior block with milder stretching. The
transition is performed in such a way as to avoid significant skewness. At the pipe centre
the cells are approximately cube-shaped ($\Delta r\approx r\Delta
\theta\approx\Delta z$). The straight-pipe mesh is then smoothly deformed to the wavy geometry by applying a streamwise-dependent radial scaling of the node coordinates so that the wall conforms as defined
in~Eq.~\eqref{r}. Mesh validity was verified by checking that element
Jacobians remain positive throughout the domain.
On the spectral grid the velocity is discretized using Lagrange polynomials of order
$N=9$ on Gauss-Lobatto-Legendre points, while pressure is represented
using a $P_{N-2}$ formulation. Non-linear convective terms are
dealiased by over-integration following the standard 3/2-rule
implemented in Nek5000.
Time advancement is performed using a semi-implicit scheme, with the
Backward Differentiation Formula (BDF) for linear terms and
extrapolation (EXT) for non-linear convective terms.
Several previous numerical studies have highlighted the importance of
adequately resolving flow near rough or modulated walls
\cite{busse2015, Chan2015, demaio}, typically recommending a minimum of
12--15 grid points per roughness element. Although the present
geometry involves smooth, large-scale periodic modulation rather than
small-scale roughness, the adopted mesh provides 24--60 radial grid
points, depending on the roughness amplitude. This ensures sufficient
resolution of near-wall gradients and recirculation features,
particularly as larger-amplitude cases occupy a greater portion of the
radial domain. The numerical setup for the different pipe geometries
under consideration is given in Table~\ref{table:1} for the bulk
Reynolds numbers $Re_b = 2000$ and $5300$. For all geometries we
consider twelve different bulk Reynolds numbers between $50$ and
$5000$.

\subsubsection{\label{sec:time}Time step and averaging time}

All simulations are performed with a
Courant-Friedrichs-Lewy (CFL) number maintained below unity to ensure
numerical stability. The timestep, expressed in wall units as $\Delta
t^+ = u_\tau^2 \Delta t / \nu$, ranges from $0.16$ to $0.489$.
After the initial transient, each case is continued for a
statistically converged averaging period, summarized in
Table~\ref{table:1}. The total averaging time, non-dimensionalized as
$T/\tau_t = T u_\tau / R_{\max}$, spans between approximately 56 and 80 eddy
turnover times, depending on the configuration. This corresponds to
about 10--32 bulk flow-through times, based on $T u_b / L_z$. These
averaging intervals are consistent with those employed in recent
high-fidelity simulations~\cite{Chan2015, Pirozzoli_2021, demaio} and
were found here to provide reliable convergence of first and second-order
statistics.

\subsubsection{\label{sec:init}Flow initialization and laminar-turbulent transition}

For the smooth-wall pipe, the flow is initialized with a
parabolic velocity profile and small-amplitude random perturbations
added to promote transition. At low bulk Reynolds numbers,
particularly in the transitional regime ($Re_b = 2500$--$4000$),
these perturbations are subject to viscous dissipation and may decay,
potentially leading to relaminarization. To prevent this, the
viscosity is temporarily reduced, equivalent to imposing a higher
Reynolds number, which allows the perturbations to grow and initiate
the transition to turbulence. An initial stabilisation phase of
approximately 5000 timesteps is used to ensure numerical stability.

For the wavy-wall pipes, the flow perturbation due to the
large-amplitude roughness naturally drives the laminar-turbulent
transition. However, to accelerate the development of a statistically
steady turbulent state, the initial velocity field is
obtained by interpolating from a fully developed smooth-wall case. A
similar stabilisation strategy is adopted, with a longer initial phase
of around 20000 timesteps used before time averaging is
initiated.

\subsubsection{\label{sec:valid}Validation}

\begin{figure}[t]
\includegraphics[width=\columnwidth]{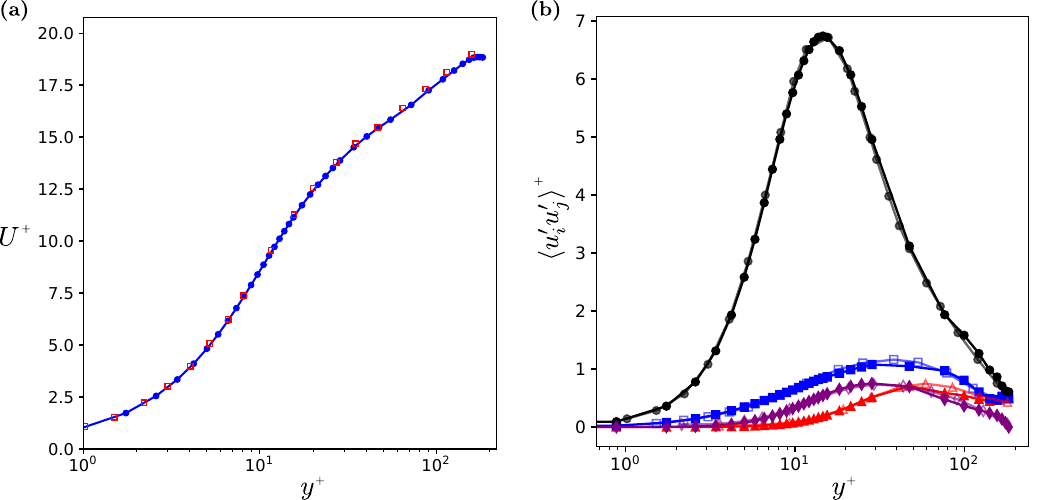}
\caption{Validation study results.
(a) Mean streamwise velocity profile in wall units, $U^{+}$ as a
function of $y^{+}$, comparing the present simulation (filled symbols)
with the reference DNS data of \citet{Pirozzoli_2021} (empty symbols)
at $Re_{\tau}=180$.
(b) Reynolds stress components normalized by the friction velocity
squared, $u_{\tau}^{2}$, as functions of $y^{+}$. Shown are the
velocity variances $\langle u_r'^2 \rangle^{+}$ ($\triangle$),
$\langle u_\theta'^2 \rangle^{+}$ ($\square$), $\langle u_z'^2
\rangle^{+}$ ($\circ$), and the Reynolds shear stress $\langle u_r'
u_z' \rangle^{+}$ ($\diamond$). Full symbols denote the present
simulation and empty symbols the reference data.}
\label{re_5300_res_2}
\end{figure}

The numerical set-up was first validated by simulating turbulent flow
in a smooth circular pipe at a bulk Reynolds number $Re_b =
5300$. The computed wall shear stress was $\tau_w = 4.6875 \times
10^{-3}$, corresponding to a friction Reynolds number $Re_\tau =
u_\tau R / \nu \approx 180$, within 1\% of the canonical value
\cite{wu_2008}. For this reference case, the near-wall mesh
resolution was $\Delta r_{\min}^+ \approx 0.09$, while the
wall-parallel spacings were $\Delta \theta^+ = 1.13$ in the azimuthal
direction and $\Delta z^+ = 2.55$ in the streamwise
direction. Accordingly, the near-wall mesh resolution is
expressed as $\Delta r^+=\Delta r u_{\tau}/\nu$, $\Delta
\theta^+=\Delta \theta u_{\tau}/\nu$, $\Delta z^+=\Delta z
u_{\tau}/\nu$, where the azimuthal spacing is based on the wall-arc
length $r\Delta\theta$.
The mesh was not tailored to this smooth configuration, but rather
designed a priori to satisfy resolution requirements under the most
demanding conditions of the study, namely the wavy pipe with the
largest geometric amplitude and highest Reynolds number.  Furthermore,
the wall-unit spacings across all rough cases remain within DNS
standards, including $\Delta r^+ \lesssim 1$, $\Delta \theta^+
\lesssim 6$, and $\Delta z^+ \lesssim 15$, even under the most severe
flow conditions as suggested for smooth-wall flows
\cite{Piomelli1993, Moser1999}. Full resolution statistics for every
case are listed in Table~\ref{table:1}.
Figure~\ref{re_5300_res_2} compares the present results for mean and
second-order statistics with those of \cite{Pirozzoli_2021}, showing
very good agreement.
Figure~\ref{re_5300_res_2}(a) presents the rescaled streamwise
velocity profile, with the mean velocity expressed in wall units as
$U^+ = u/u_{\tau}$ and the wall-normal coordinate as $y^+ = y
u_{\tau}/\nu$. Figure~\ref{re_5300_res_2}(b) shows the turbulence
statistics, including the velocity variances $\langle u_r'^2 \rangle^+$,
$\langle u_{\theta}'^2 \rangle^+$, $\langle u_z'^2 \rangle^+$, and the
Reynolds shear stress $\langle u_z' u_r' \rangle^+$.

\section{\label{sec:results}Results}

\begin{figure}[b]
\includegraphics[width=100mm]{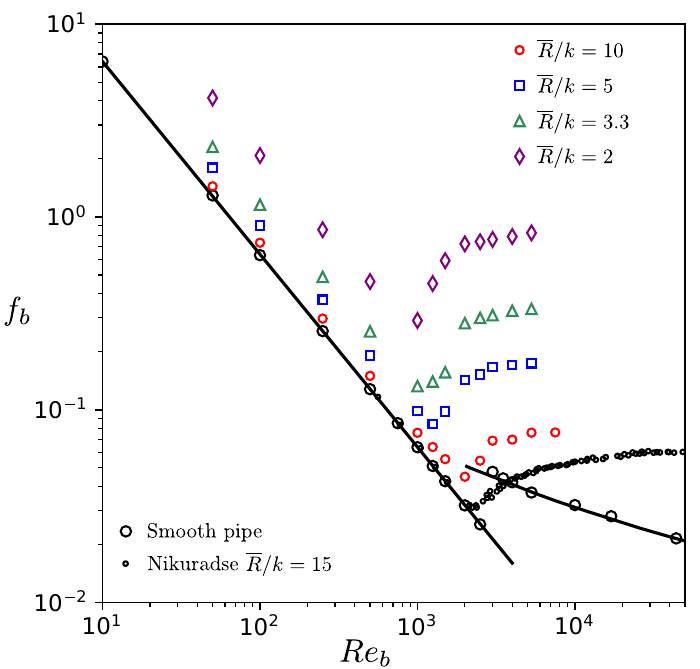}
\caption{Friction factor ($f_b$) versus bulk Reynolds number
($Re_b$), evaluated with the maximum radius ($R_\mathrm{max}$). As a reference, results for the smooth pipe are reported together with the data of \cite{nikuradse1933} for ${\overline{R}/k} = 15$.}
\label{f_1}
\end{figure}

The results are organized according to flow regime, using the friction
factor as the primary diagnostic to characterize the influence of
surface waviness and flow inertia. This choice reflects the central
role of global pressure losses in assessing the hydraulic impact of
geometric roughness, while allowing local flow features to be
interpreted within a consistent framework.



Figure~\ref{f_1} shows the friction factors obtained from the
present simulations across the explored bulk Reynolds number range. The
results are shown together with the smooth-pipe prediction and
Nikuradse's classical data \cite{nikuradse1933} for the highest
available relative roughness ($\overline{R}/k_s = 15$), providing a
reference framework against which the different flow regimes and
roughness effects discussed in the following sections can be
interpreted. Several systematic departures from the classical behavior are
immediately apparent. In the laminar regime, the friction factor does
not collapse onto the Hagen-Poiseuille prediction when expressed in
terms of the bulk Reynolds number $Re_b$, but instead exhibits an
upward offset that increases with wall waviness. As the Reynolds
number is increased, the transition away from laminar flow occurs
progressively earlier than in a smooth pipe and displays a clear
dependence on the geometric modulation, raising the question of how
wall waviness alters flow stability and the transition mechanism
itself. At sufficiently high Reynolds numbers, the friction factor
becomes weakly dependent on $Re_b$ and instead organizes according
to roughness level, signaling the emergence of a fully rough
regime at friction values significantly larger than for classical
rough pipes. 
These observations motivate the regime-based analysis that
follows, in which each of these questions is examined and quantified
in turn.

\subsection{\label{sec:lf}Laminar flow}

\begin{figure}[b]
\includegraphics[width=\columnwidth]{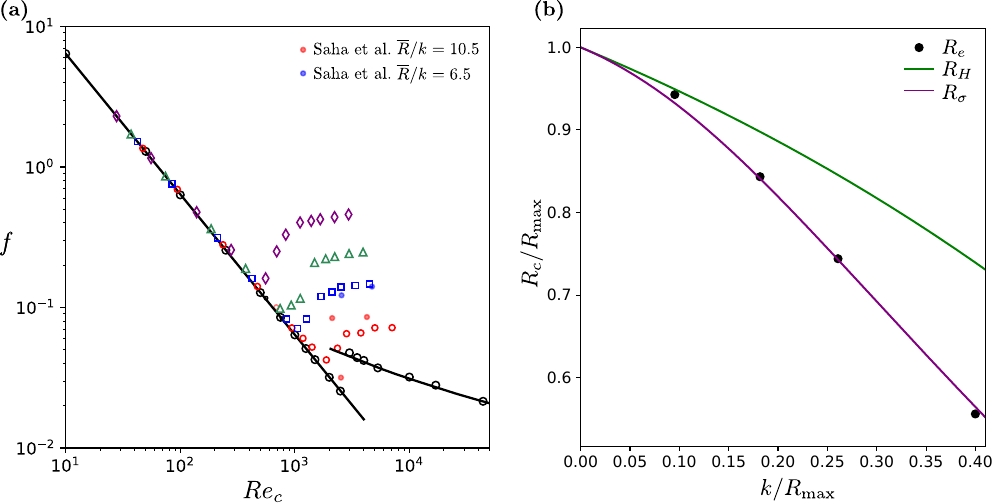}
\caption{(a) Variation of friction factor with the effective hydraulic Reynolds number ($Re$), based on the effective radius ($R_h$) evaluated from the laminar simulations ($R_h = \sqrt{8\mu u_b/G}$).
The solid line denotes the Hagen-Poiseuille law ($f = 64/Re$) and
the Prandtl friction laws for laminar and turbulent flow. As a
reference, results for the smooth pipe are reported together with the
data of \cite{nikuradse1933} for ${\overline{R}/k} = 15$, and
the results of the investigation by \citet{saha_2015} for the
geometries with ${\overline{R}/k} = 10.5$ and 6.5. (b) Non-dimensional
radius ratios $R_c/R_{\max}$ as a function of the relative roughness
$k/R_{\max}$ for the different radius definitions considered. The
flow-derived radius $R_h$, obtained from laminar simulations, is
included as a reference.}
\label{f_2}
\end{figure}

We begin the regime-based analysis with the laminar flow region, where
the friction factor provides a direct and unambiguous measure of
hydraulic resistance. In this regime, turbulence is absent and energy
dissipation is controlled primarily by viscous effects and geometric
constraints, making it an ideal starting point for isolating the
influence of wall waviness on flow resistance. Figure~\ref{f_1}
therefore serves as a reference framework for assessing how deviations
from classical laminar behavior arise as a consequence of geometric
distortion rather than inertial or turbulent mechanisms.

\subsubsection{\label{sec:hydrad}Effective hydraulic radius}

In the laminar regime ($Re_b\lesssim 2000$), Nikuradse's data
collapse onto the theoretical Hagen-Poiseuille prediction, $f_b =
64/Re_b$, regardless of the roughness level, which implies that geometric
roughness does not alter significantly the laminar friction in
those experiments. By contrast, our numerically obtained laminar
friction factors exceed the
theoretical value by 5--80\%. Two effects explain the
discrepancy. First, defining the Reynolds number using the maximum  radius
$Re_b=2R_{\max}u_b/\nu$ overestimates the effective cross-sectional
area available to the flow. Second, and more fundamentally, even when
the mean radius is adopted, the large relative roughness of the wavy
geometry reduces the flow conducting cross section and can lead to
secondary flow structures in the divergent pipe
regions that raise the flow resistance. 

For pipes with strong wall modulations, the effective radius that
collapses the data for the friction factor onto the curve $f =
64/Re$ is in fact a hydraulic quantity just like the effective
sandgrain roughness discussed in Sec.~\ref{sec::ks}.
Therefore, the effective hydraulic radius $R_e$ is defined in the spirit of
an effective conductance as the radius of an
equivalent straight pipe that has the same laminar flow resistance as
the wavy pipe. The Hagen-Poiseuille law thus gives
\begin{align}
R_e = \sqrt{\frac{8 \mu u_b}{G}}.
\label{hydraulic_r}
\end{align}
In the following, we define the wall friction, Reynolds number and
friction factor in terms of $R_e$, that is, we set $R_c = R_e$ in
Eqs.~\eqref{eq:tauw}, \eqref{eq:Retau} and \eqref{f_eq}.
With these definitions, the laminar data of the
friction factor aligns with $f = 64/Re$, and therefore
embeds the influence of the distorted wall shape in the length scale
itself. As shown in Figure~\ref{f_2}(a) this rescaling collapses
the laminar friction factor data onto the theoretical line, using the
adjusted Reynolds number $Re$.
The figure also includes the
DNS results of \citet{saha_2015}, for two
comparable roughness levels (${\overline{R}/k} = 10.5$ and
6.5), which display the same qualitative behavior. However,
quantitative differences remain, which can be traced back to the fact
that these authors use the mean radius as the characteristic length scale.
The choice of the characteristic length scale does not only
affect friction factor in the laminar regime, but also in the
transitional and turbulent regimes.
The effective hydraulic radius defined by Eq.~\eqref{hydraulic_r} is
in the first place a hydraulic and not a geometric quantity.
In order for the friction factor to be predictive for
the mean flow behavior, it is desirable to relate $R_e$ to the
geometric characteristic of the wall fluctuations.

For creeping flow through a wavy pipe, lubrication theory
provides a systematic reduction of the governing equations under the
assumptions of slow axial variation and negligible inertia, leading to
generalized expressions for the friction factor in periodically
constricted tubes. Such analyses have been applied to a range of
geometries, including sinusoidally varying pipes and non-sinusoidal
converging-diverging configurations. Several studies \cite{chow_1972,
deiber_1979, Sisavath_2001} derived lubrication-based models for
laminar flow through tubes with sinusoidal diameter variations,
showing that the leading-order viscous resistance is governed by a
harmonic-type averaging of the local radius. Similar conclusions were
reached in both experimental and numerical investigations of
periodically constricted tubes, which consistently reported an
increase in laminar resistance with increasing geometric modulation,
up to the onset of flow separation. Thus, a characteristic radius can
be defined by the following harmonic power-average,
\begin{equation}
    R_H = \left[ \frac{1}{\lambda} \int_0^\lambda
    \frac{1}{R^4(z)} \, dz \right]^{-1/4}.
\end{equation}
It represents the natural effective length scale emerging from
lubrication theory for pipes with small wall fluctuations, that is
$\overline R/\lambda \ll 1$. Unlike the arithmetic mean, it assigns greater
weight to the narrowest sections of the geometry, which dominate
viscous dissipation in laminar and creeping flows.
This weighting makes it particularly suitable for periodically
corrugated geometries such as the present wavy pipe, where
constrictions strongly influence hydraulic
resistance. In this sense, it acts as a one-dimensional geometric
simplification of the full three-dimensional geometry, compressing
axial variations into a single representative radius.

\begin{figure}[b]
\includegraphics[width=\columnwidth]{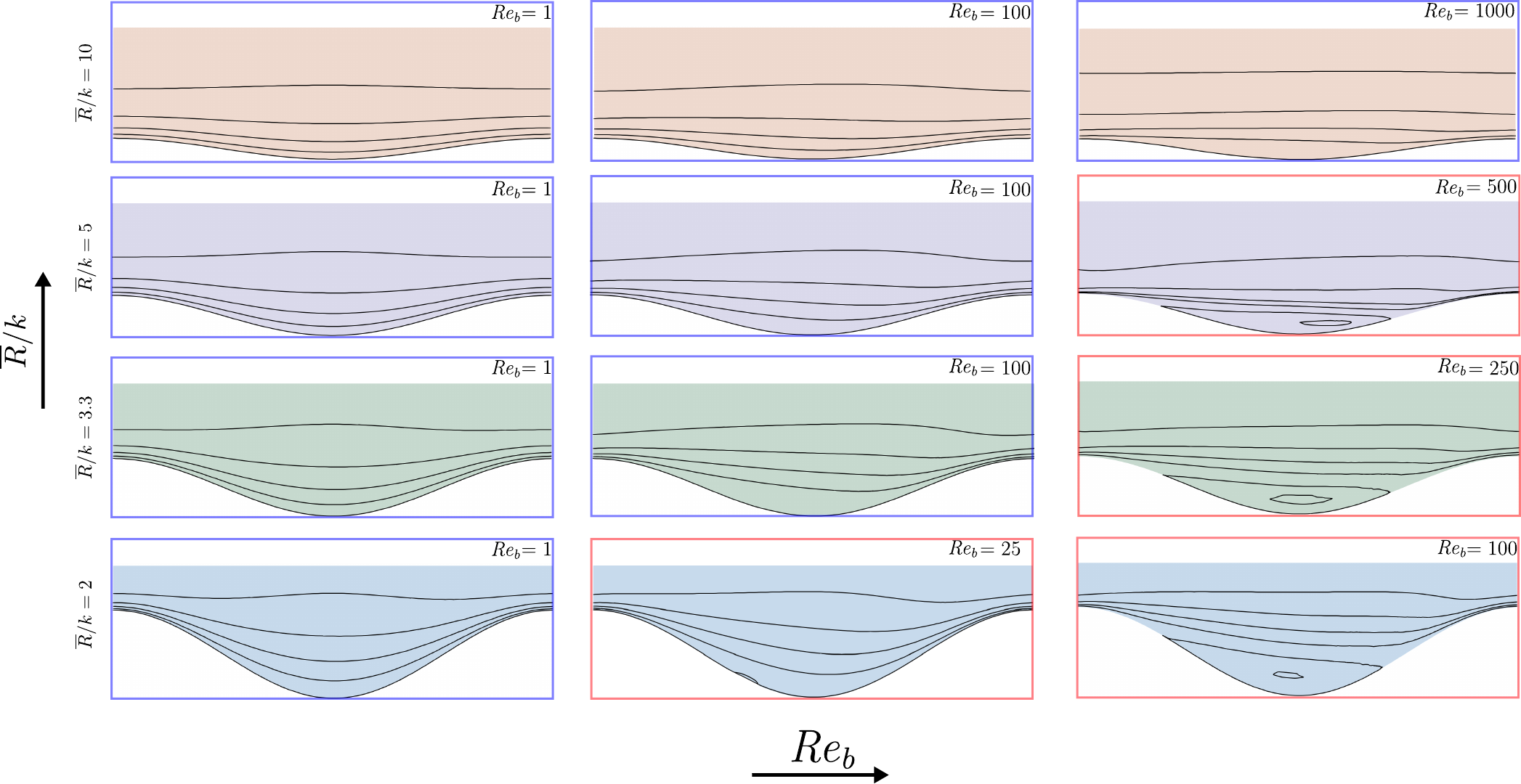}
\caption{Time-averaged isolines of the main component of the
velocity field in the lower half of the pipe cross-section,
highlighting flow separation and recirculation near the diverging
region of the wavy wall. A set of 3 cases for different relative bulk
Reynolds numbers ($Re_b$) are reported for each pipe configuration. Reynolds
values were selected non-uniformly to highlight the earliest Reynolds
number at which recirculation first appears in each case.}
\label{isoline}
\end{figure}

Here, we introduce a slope-corrected radius $R_\sigma$ based on the
harmonic power-average in order to approximate $R_e$. This quantity
augments the harmonic radius $\langle R \rangle_{\mathrm{harm}}$
with a correction based on a dimensionless measure of wall slope,
designed to capture the additional dissipation associated with
geometry-induced velocity gradients and incipient separation.
Specifically, we define
\begin{equation}
R_\sigma
= R_H
\left(1 + C \sigma\right)^{-1/4},
\label{slope_mm}
\end{equation}
where
\begin{equation}
\sigma
=
\frac{1}{\lambda}
\int_0^\lambda
\left[
\frac{1}{R(z)}\frac{dR(z)}{dz}
\right]^2
\,dz
\end{equation}
quantifies the mean squared wall slope over one wavelength. The
multiplicative form preserves the correct weak-corrugation limit,
$\sigma\rightarrow 0$, for which $R_\sigma\rightarrow
R_H$, while introducing an additional resistance
contribution associated with wall-induced cross-stream motion. The
constant $C$ is a fitting parameter that is expected to be of order
one. For 
stronger corrugations, finite wall slopes force cross-stream 
velocities of order $u_r \sim u_z \, dR/dz$, generating additional 
dissipation beyond the lubrication approximation that $\sigma$ 
is designed to capture. 
Figure~\ref{f_2}(b) shows the dependence of the effective hydraulic
radius $R_e$ on the wall-to-wall amplitude $k$, and the performance of
different definitions for the characteristic radius. The effective
hydraulic radius decreases steeply with increasing wall fluctuations,
much faster than the arithmetic and harmonic power-average.
Nevertheless, $R_H$ shows excellent agreement
with $R_e$ for weak corrugations (${\overline{R}/k}\gtrsim
5$). However, for stronger waviness, $R_H$
systematically exceeds the theoretical value, indicating that
geometric constriction alone cannot explain the observed
resistance. 
The correlation~\eqref{slope_mm} for $C=1.5$ provides an
excellent fit to the numerical data, yielding residual errors of order
$1\%$.

\subsubsection{\label{sec:flowstruct}Flow structure}

The breakdown of purely geometric scaling shown earlier is directly
tied to changes in the laminar flow structure. In particular,
boundary layer separation and recirculation introduce dissipation mechanisms that
cannot be captured by one-dimensional geometric radius
definitions. These effects are illustrated in Figure~\ref{isoline},
which shows time-averaged streamlines of the streamwise velocity in
the lower half of the pipe, revealing recirculation zones forming in
the diverging regions of the wavy wall.
Panels are ordered by decreasing $\overline{R}/k$ (top to bottom) and increasing $Re_b$
(left to right). For each case, the first appearance of a
recirculation bubble is outlined in red, which provides an upper bound
for the separation threshold. Recirculation may still occur at
intermediate Reynolds numbers not sampled here. The discrete values
chosen span a broad range, with more values at low $Re_b$ and near
the onset of separation, while keeping computational cost manageable.

Boundary layer separation and the formation of a recirculation bubble
are related to the occurrence of adverse pressure gradients for flow
over curved surfaces. In wavy pipes, the succession of diverging and
converging sections leads to favorable and adverse pressure
gradients. At the diverging sections of the pipe, the flow
velocity decreases, which imprints an adverse pressure gradient on the
boundary layer, causing deceleration and eventually flow reversal
\cite{schlichting2016boundary} located at the downstream side of the
constriction before the location of the maximum diameter. The
likelihood of flow separation increases for large amplitude to
wavelength ratios \cite{schlichting2016boundary}.
In fact, for the smoothest geometry ($\overline{R}/k = 10$), the flow remains
symmetric and attached up to $Re_b = 1000$. This indicates that
viscous diffusion dominates and that mild curvature alone does not
induce separation.

At intermediate waviness ($\overline{R}/k = 5$ and
$3.3$), the velocity field becomes progressively distorted as inertia
grows. Separation first appears at $Re_b = 500$ ($Re \approx 420$)
for $\overline{R}/k = 5$ and at $Re_b = 250$ ($Re \approx 186$)
for $\overline{R}/k = 3.3$. These two cases show that moderate
curvature can combine with non-linear convection to trigger flow
reversal.
The roughest geometry ($\overline{R}/k = 2$) exhibits the
earliest onset. No bubble is present at $Re_b = 1$, but a small
steady vortex forms at $Re_b = 25$ and grows significantly by
$Re_b = 100$.
This behavior is consistent with previous 2D
studies. \citet{nishimura1984} and \citet{wang1995} reported early
recirculation for sinusoidal channels
corresponding to $\overline{R}/k \approx 0.93$, with separation
appearing at $Re_b \approx 15$ to $20$. Similarly,
\cite{al-daamee_2024} observed secondary flows at even lower values,
$Re \approx 5$. Although their configurations differ and are
strictly two dimensional, these results support the trend seen in the
present pipe flow. Increasing waviness promotes separation at
progressively lower Reynolds numbers. The location and extent of the
recirculation region depend on both $Re_b$ and $\overline{R}/k$,
showing that the geometry imposes a preferred site for instability.
This is in line with the theory of boundary layer separation
\cite{schlichting2016boundary} that places the recirculation region
at the diverging section before the trough, where adverse pressure
gradients are largest.

Thus, in all cases where separation occurs, reversed flow initiates near the
lower wall on the downstream side of the constriction. As
$\overline{R}/k$ decreases, the separation point moves upstream and
the recirculation zone expands, shifting the instability closer to the
throat. At higher $Re_b$, the vortex intensifies and its centre
moves downstream.
These structural changes directly affect pressure
losses. Recirculation increases shear and energy dissipation through
separation and reattachment, even under nominally laminar
conditions. The early appearance of recirculation therefore helps
explain the elevated friction factors and earlier transition to turbulence observed in wavy pipes, as discussed in the next section.

\subsection{\label{sec:lt}Transitional flow}

This section examines the breakdown of laminar flow considered in the wavy-pipe
geometries. Transition is characterized using global flow
metrics and temporal flow behavior, with emphasis on changes in
friction factor and the persistence of velocity fluctuations as the
Reynolds number increases.

\begin{figure}[t]
\includegraphics[width=100mm]{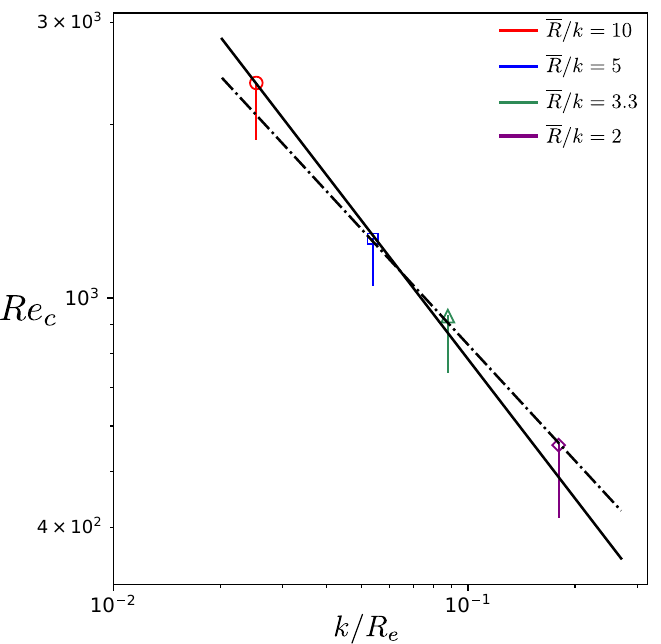}
\caption{Scaled roughness amplitude $k/R_e$ as a function of the effective Reynolds number $Re=\frac{2R_eu_b}{\nu}$ for all four geometries. The markers indicate the lowest Reynolds number at which sustained turbulence was observed. Vertical lines indicate the uncertainty interval within which the  transition threshold lies. The solid line shows a reference scaling proportional to $Re^{-3/2}$, the dashed line $Re^{-4/5}$.}
\label{trans}
\end{figure}

\subsubsection{\label{sec:critRe}Critical Reynolds number}

Transition in smooth pipes is a classical subcritical problem. The
laminar base state is linearly stable, and turbulence arises only from
finite-amplitude disturbances. As a result, transition occurs through
intermittent puffs and slugs whose appearance depends sensitively on
disturbance amplitude and flow history
\cite{avila_2011, Barkley_2016, cerbus_2018, avila_2023}.
Experiments using controlled perturbations \cite{NISHI_2008}
demonstrate that geometric forcing provides a deterministic
disturbance that determines the Reynolds number at which sustained
non-laminar flow appears. \citet{grossmann_2000} estimates the
critical amplitude required for positive feedback to scale as
$Re^{-\gamma}$ with $\gamma$ of order 1. \citet{chapman_2002}
provides a detailed analysis of the steps leading to a positive
feedback loop in order to derive the scaling of the amplitude of the
initial perturbation with $Re$. He discusses two routes to turbulence
depending on the orientation of the initial perturbation. For a
streamwise perturbation, he obtains the
scaling $v_0 \propto Re^{-3/2}$ of the perturbation amplitude $v_0$
with $Re$. For an oblique orientation, he
derives $v_0 \propto Re^{-5/4}$. The experiments of
\citet{peixinho2007finite} find power-law scalings of the critical
perturbation amplitude with $\gamma = 1.3$ and $1.5$ depending on the orientation
of the perturbation. Finite-amplitude (bypass) transitions in linearly stable shear flows
can be assessed using the non-normality of the linearized
Navier-Stokes operator for laminar base flow. The solution of the
linearized system gives a transient linear increase of an initial
velocity perturbation of order $Re$ before it decays. The non-linear
term in the Navier-Stokes equation can channel the perturbation from
the decaying back into the growing part of the solution, which leads to a
positive feedback loop \cite{trefethen1993hydrodynamic,
brandt2014lift}. These dynamics can be mapped onto predator-prey type
models \cite{shih2016ecological,wang2022stochastic} to quantify the
competition between turbulent (prey) and laminar (predator) flow
modes. The decay and proliferation of turbulent puffs
described by this type of dynamics gives strong indications that the laminar
turbulent transition for smooth pipes belongs to the directed
percolation universality class \cite{shih2016ecological, avila_2023}.  

This is very different for wavy pipes. Here, the imposed disturbance is steady, spatially
distributed, and intrinsic to the geometry. As discussed
in the previous section, under laminar conditions, a vortex is created
when the adverse pressure gradient at the diverging pipe section leads
to boundary layer separation. The creation of a recirculation zone can
be seen as a precursor for the early onset of transitional flow
\cite{cotrell2008instability}. The onset of
non-laminar flow is identified here by the appearance of persistent
velocity fluctuations and by deviations of the friction factor from
the laminar trend. Because simulations run over a finite time window,
the reported transition Reynolds numbers should be interpreted as
upper bounds. Figure~\ref{trans} summarizes the transition thresholds
by plotting the critical Reynolds number $Re_c$ versus the normalized wall
amplitude. For each roughness level, the symbol marks the lowest
Reynolds number for which sustained turbulence was observed. The vertical line
extends downward to the highest Reynolds number at which the flow
remained laminar, indicating that the true transition threshold lies
somewhere within this interval. 
The data suggest an approximate scaling of $Re_c$ as 
\begin{align}
\label{eq:k_scaling}
Re_c \sim \left(\frac{k}{R_e} \right)^{-1/\gamma},
\end{align}
with $5/4 \leq \gamma \leq 3/2$. 

The scaling of Eq.~\eqref{eq:k_scaling} establishes that the transition is governed by a
finite-amplitude mechanism consistent with that observed for the
transition in smooth pipes, as discussed above. It does not, however, explain why the wavy
geometry is so much more susceptible to turbulence than a smooth pipe. The reason is a change
in the stability character of the base flow caused by the recirculation. In a
smooth pipe the laminar profile is parabolic, so its second radial derivative
$U''$ is of constant sign \footnote{For brevity, we use the notation
$U' = \partial U(r)/\partial r$ to denote the radial derivative}. The
profile has no inflection point and therefore satisfies neither the Rayleigh nor
Fj$\o$rtoft criteria~\cite{drazin2004} for instability. This is the classical reason why
Hagen--Poiseuille flow is linearly stable at all Reynolds numbers.
The separated profile in the diverging sections of the wavy pipe is qualitatively different.
Fig.~\ref{isoline} shows that the streamwise velocity across the expansion
section toward $R_{\max}$ (see the case with $\bar{R}/k=2$ at
$Re_b=100$ in a steady laminar state) is negative in a thin layer at the wall. The profile is thus
non-monotonic and its curvature reverses. An inflection point $y_s$ appears in the
shear layer between the reversed near-wall fluid and the forward core, so that
both criteria, $U(y_s)'' = 0$ (Rayleigh) and $U''[U-U(y_s)]<0$ (Fj$\o$rtoft), are satisfied close to
the wall. Wall waviness may play a role similar to finite amplitude
perturnations in smooth pipes. 
More than this, it makes the base flow itself inflectional and opens an
inviscid instability route that is absent in the straight geometry.
This route is not only permitted but operative. This picture suggests
that the gradual transition to turbulence with increasing $Re$ comprises two stages, which
describe a supercritical road to turbulence. First, the formation of a laminar
vortex in the expanding pipe sections, leading to a linearly unstable
laminar base flow. Second, the transition to turbulence from
quasi-periodic to chaotic flow, potentially through a
Kelvin-Helmholtz-type instability, as observed for flow in wavy
channels \cite{rivera2013global,harikrishnan2021flow}.  

The linear stability of wavy pipes with small wall fluctuations has
been discussed by~\citet{cotrell2008instability} for $k/\overline R \leq 0.2$ and
wavelength $\lambda = \overline R$. These authors point out that the
corrugated base flow is linearly stable for wall amplitudes below a
threshold value. For wall corrugations above the treshold, flow becomes linearly 
unstable once a vortex forms in the diverging section. They track the
destabilization to the coupling between the radial gradient of the
base axial velocity and the radial disturbance velocity, which
is consistent with the inflectional mechanism described above, and the
creation of a Kelvin-Helmholtz-type instability. 
Also, they find that the critical Reynolds number depends on the
amplitude as $Re_c \sim k^{-1}$ consistent with the results shown in
Figure \ref{trans} and the analysis of \citet{grossmann_2000}.

Taken together, the inflectional character of the base flow and the
linear stability fix the role of the geometry. 
For the geometries under consideration here with corrugations
$k/\overline R \geq 0.1$ and fixed wavelength $\lambda = 2 R_{\max}$,
the transition occurs at $Re$ between $500$ and $1000$ consistent with
the experimental findings of \citet{deiber_1979} and \citet{nishimura_2003}. The critical
Reynolds number in general depends on the relative wall amplitude and wave
number, which vary here simultaneously because we keep the maximum
radius and wavelength fixed. The observed dependence of $Re_c$ on $k$
therefore represents an empirical upper bound for transition in the
family of wavy geometries under consideration here, rather than a
universal transition law. A systematic investigation of the impact of
wavenumber and amplitude is beyond the scope of this
study. Nevertheless, it provides clear quantitative
evidence of the onset of turbulence at significantly lower $Re$ than
for smooth and classical rough pipes, in accordance with observations
in karst conduits \cite{worthington2017}.

\begin{figure}[b]
\includegraphics[width=\columnwidth]{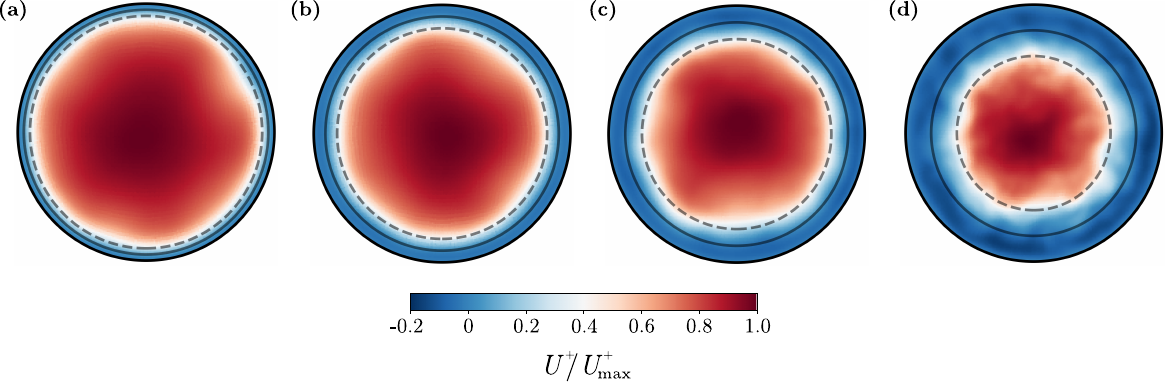}
\caption{Contours of mean streamwise velocity ($U^+/U_{max}^+$) for the different
roughness level for the case $Re_\tau\approx120$ extracted at the
correspondence of the divergence $R=R_{\max}$. The solid line
represents the mean radius ($\overline{R}$) and the dashed line
depicts the plane of the peak of the roughness ($R=R_{\min}$).}
\label{contours_ks}
\end{figure}

\subsubsection{\label{sec:meanvel_trans}Mean velocity}

To further characterize the transitional regime, the flow statistics
are analysed in more detail at selected operating conditions. In order
to facilitate comparison across different geometries, cases are
examined at matched friction Reynolds number,
$Re_\tau\approx120$. Since the friction Reynolds number varies
significantly between smooth and rough cases at the same $Re_b$,
matching $Re_\tau$ ensures a physically meaningful comparison of
turbulent flow statistics between rough and smooth-wall cases,
following standard practice in rough-wall turbulence studies
\cite{chung2021}. This choice allows differences in the mean flow
and fluctuation statistics to be attributed primarily to geometric
effects rather than variations in wall forcing. The following analysis
focuses first on mean velocity statistics, including local and
cross-sectionally averaged profiles. Second-order statistics are
provided in Appendix \ref{app}. 

Figure~\ref{contours_ks} presents contours of the mean
streamwise velocity $U^+$, extracted at the axial location
corresponding to the maximum cross-sectional area, for
increasing levels of surface roughness. In panel (a), corresponding to
the smallest roughness amplitude ${\overline{R}/k}=10$, the
flow preserves a near-parabolic profile, although slight deviations
from symmetry are already visible. As roughness increases from panels
(b) through (d), the velocity distribution becomes progressively
flatter and more distorted. A significant drop in core velocity
appears, accompanied by steeper gradients near the wall. Indications
of recirculation appear in panel (b), although they remain weak and
confined to narrow regions. In panel (c), these zones become more
prominent, with distinct regions of reversed flow visible particularly
in the region between the mean and maximum radius. By panel (d),
recirculation is clearly established and occupies a substantial
portion of the near-wall region. The solid and dashed circles in each plot mark the mean radius
$\overline{R}$ and the minimum radius $R_{\min}$, respectively, helping
to visualize the relative displacement of high- and low-velocity
regions within the perturbed geometry. As roughness increases, the
near-wall region becomes dominated by reverse and low-speed flow,
forming a multilayered structure that departs from classical wall
scaling. This structure consists of a reversed near-wall zone, a
velocity-deficit region, and a high-shear interface with the bulk
flow.

\begin{figure}[b]
\includegraphics[width=100mm]{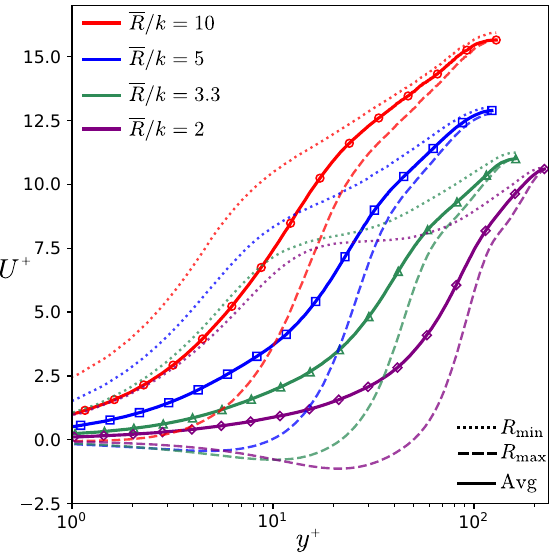}
\caption{Inner-scaled mean velocity profiles $U^+$ versus wall-normal distance
$y^+$ in wavy pipes with varying roughness configurations at
$Re_\tau \approx 120$.
Profiles are shown at the cross-sections corresponding to the maximum and minimum pipe radius ($R_{\max}$ and $R_{\min}$), together with the streamwise-averaged profile.}
\label{ks_2000_2}
\end{figure}

Figure~\ref{ks_2000_2} shows the inner-scaled mean streamwise velocity profiles as a function of wall-normal distance. The profiles are extracted at cross-sections corresponding to the minimum radius $R_{\min}$ and maximum radius $R_{\max}$, and are compared with the streamwise-averaged profile. The results reveal a strong spatial modulation of the mean flow induced by the wavy geometry.

In the constricted sections ($R_{\min}$), the flow accelerates under a favorable pressure gradient, leading to elevated values of $U^+$ relative to the streamwise average. In contrast, at $R_{\max}$ the adverse pressure gradient in the expansion region produces a pronounced velocity deficit and, for $\overline{R}/k \ge 5$, a near-wall flow reversal consistent with the separation and reattachment observed in Fig.~\ref{contours_ks}. As roughness increases, the $R_{\max}$ and streamwise-averaged profiles progressively diverge, reflecting the growing influence of separation.
Despite these differences, the profiles collapse up to $y^+ \approx 6$ for all roughness levels, indicating that the immediate viscous sublayer remains governed by friction scaling. Beyond this region, the recirculation zones thicken the low-speed region and shift the profiles towards the pipe centre. Downstream of reattachment, the flow re-accelerates and partially recovers a common trend, suggesting that the geometric perturbation is spatially localized rather than globally disruptive.
Relative to the smooth-pipe reference, all rough cases exhibit a downward shift of the inner-scaled mean velocity profile.

\subsection{\label{sec:ltu}Turbulent flow}

This section examines the flow behavior at the highest Reynolds numbers
considered in this study, where velocity fluctuations are sustained and
the flow is turbulent over the wavy wall geometries. In this regime,
geometric measures such as effective diameters become insufficient to
describe the drag response, and velocity-based metrics provide a more
appropriate framework for characterizing roughness effects. The
analysis therefore focuses on the roughness function and the equivalent
sand-grain roughness as primary descriptors of the turbulent response.
The implications of this parametrization are subsequently examined
through mean velocity profiles. Second-order statistics are reported
in Appendix \ref{app}.

\begin{figure}[b]
\includegraphics[width=\columnwidth]{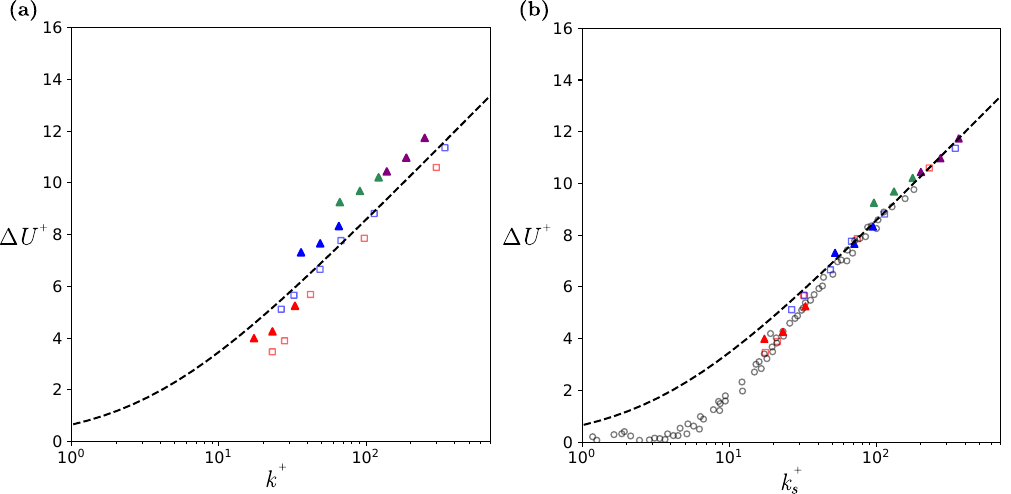}
\caption{(a) Variation of roughness function ($\Delta U^+$) with 
inner-scaled roughness height ($k^+ = k u_\tau/\nu$). (b) Same 
roughness function plotted against the equivalent sand-grain 
roughness height ($k_s^+$). As a reference the plots contain empty 
black circles that represent the data from Nikuradse's experiment. 
The squares denote data from De Maio \textit{et al.}~\cite{demaio}. 
The dashed line represents the Colebrook relation~\cite{Colebrook1939}.}
\label{re_study_1}
\end{figure}

\subsubsection{\label{sec::ks}Equivalent sand-grain roughness}

As discussed previously, the traditional use of effective geometric
diameters may underpredict the pressure loss in rough pipes, especially in
regimes where flow separation begins but turbulence has not yet fully
developed. This stresses the limitations of empirical models that rely
solely on geometric corrections to predict pressure losses in rough or
undulating conduits. The effective hydraulic radius $R_e$ is a hydrodynamic concept
that relates flow rate and pressure drop in the laminar
regime. In the same vein, the equivalent sand-grain roughness is a hydrodynamic
concept that characterizes the flow in the fully
rough regime \cite{jimenez2004}. The equivalent sandgrain roughness $k_s$
determines the asymptotic value of the friction factor $f(Re)$ in
the fully rough regime. It
can be defined equivalently by inverting the roughness function given by
Eq.~\eqref{colebrook},
\begin{align}
\label{eq:ks}
k_s = \frac{\exp\left(\kappa \Delta U^+\right) -1}{0.3},
\end{align}
where $\Delta U^+$ is the velocity deficit in the logarithmic region defined in
Eq.~\eqref{eq:deltaU} with respect to a smooth profile at the same
$Re_\tau$ obtained from the DNS. As
outlined in Sec.~\ref{sec:loglaw}, the wall roughness moves the
flow toward the core region of the conduit such that the logarithmic
law given by Eq.~\eqref{eq:Urough} is expected to hold as a function
of $Y = y - \ell$ with $\ell$ the wall offset defined in Appendix~\ref{app:offset}. Thus, $\Delta U^+$ is determined from the DNS
data in the wavy pipes subject to this shift. The  roughness function is computed as the vertical offset between the 
smooth and rough-wall velocity profiles in the logarithmic region. 
The smooth reference is taken from simulations of a smooth pipe with same radius of $R_{max}$ at comparable friction Reynolds numbers $Re_{\tau}$, allowing consistent comparison despite the non identical wall stress.

In Figure~\ref{re_study_1}(a), the roughness function $\Delta U^+$
obtained in this way is shown as a function of the
roughness Reynolds number $k^+=ku_\tau/\nu$ for different friction
Reynolds numbers ($Re_\tau = 180$, $240$, and $325$) and roughness levels.
For reference, data from the rough-pipe simulations of \citet{demaio} are also included.
The roughness function $\Delta U^+$ increases from approximately 4 for
the mildest roughness to about 8 for the largest amplitude considered,
indicating progressively stronger frictional resistance.
The values of $\Delta U^+$ and $k^+$ obtained for
the wavy-pipe cases collapse reasonably well onto the Colebrook
relation, confirming that even periodic and deterministic wall
geometries produce drag increases comparable to those observed over
irregular rough surfaces. For instance, the case with $\overline{R}/k
= 2$ and $Re_\tau \approx 325$ yields $\Delta U^+ \approx 11.2$,
corresponding to $k^+ \approx 250$, which lies near the upper limit of
typical engineering roughness values. \citet{Flack2014} reported that
wall Reynolds numbers for most practical rough surfaces fall within
$k_s^+ < 800$. The trend observed among the different geometries is
consistent with the results of \citet{demaio}, with roughness
functions exceeding $\Delta U^+ = 7$ for the cases with
$\overline{R}/k = 5, 3.3$ and $2$, indicating that
these flows fall within the strongly fully rough regime. In contrast,
for the less wavy geometries with $\overline R/k = 10$, the roughness function remains below
this threshold, corresponding to laminar or transitional roughness
behavior within the range of flow conditions investigated, where
viscous effects still exert a dominant influence.

An equivalent sand-grain roughness height $k_s^+$ was estimated by
fitting the Colebrook relation to the DNS data of the roughness
function for $\overline R/k = 2, 3.3$ and $5$.
The value $k_s^+ \approx 1.5k^+$ fits all the data, indicating that the
effective hydraulic roughness is of the order of the
geometric amplitude of the corrugation. For the case $\overline R/k = 10$,
we find $k_s = k$. The resulting comparison,
reported in Fig.~\ref{re_study_1}(b), includes data from
\citet{demaio} and the classical experiments of Nikuradse as reference
cases. A nearly perfect collapse onto the Colebrook curve is observed
for most geometries; only the mildest roughness case shows a slight
deviation, which may be attributed to the flow not having reached the
fully rough regime. This distinction highlights the inherent difference between the weakly corrugated case, where viscous effects and classical roughness scaling remain relevant, and the strongly modulated geometries, where form-induced drag dominates and the geometric amplitude $k$ provides a robust hydrodynamic length scale. The obtained value, $k_s^+ \approx 1.5k^+$,
reinforces the choice of the geometric amplitude $k$ as the relevant
length scale for these wavy-wall geometries, consistent with the
self-similar behavior observed for the mean velocity reported in the
next section.

To further examine the relationship between surface geometry and flow
response, the equivalent roughness heights $k_{s}^+$ from Eq.~\eqref{eq:ks}
are compared with geometric and empirical estimates, as reported in
Table~\ref{tab:ks_methods}. The analysis was
performed for the two geometries at $\overline{R}/k = 3.3$ and
$\overline{R}/k = 2$, corresponding to the reference values of $k_{s}^+ = 176.1$
and $361.9$, respectively, considering the highest Reynolds number
($Re_\tau\approx325$).
Among the tested models, the correlation of \citet{Chan2015} provides the
most accurate prediction. Models developed for irregular
roughness \cite{napoli2008, Flack2020} tend to deviate
more strongly. In line with our observation that the sandgrain
roughness is of the order of the peak-to-peak distance,
\cite{saha_2015} propose to use $k_s = 2 k$. In summary, our results show that the
geometric amplitude $k$ as the primary length scale controlling the
drag response for these smooth, directly sets the sandgrain roughness
for periodic wall modulations.

\begin{table}[b]
  \caption{Estimated equivalent sand-grain roughness heights $k_s$ for the
sinusoidal-walled pipe cases at $Re_\tau = 325$. Each correlation is evaluated
using the geometric peak-to-peak roughness $k$ and wavelength $\lambda = 1$.
The resulting $k_s^+$ (pred.) values predicted from the various empirical correlations
are compared with the values $k_{s}^+$ (ref.) obtained from the logarithmic-law offset
derived from the simulations.
The percentage error is computed relative to the reference $k_{s}^+$ (ref.)
for each case. The two cases
correspond to (1) $\overline{R}/k = 3.3$ and (2) $\overline{R}/k = 2$.
The references refer to Na2008 = \citet{napoli2008}, Ch2015 = \citet{Chan2015}, Sa2015
= \citet{saha_2015}, Fo2017 = \citet{Forooghi2017}, FS2020 = \citet{Flack2020}.}
  \label{tab:ks_methods}
\begin{ruledtabular}
\begin{tabular}{lclccc}
Case & Ref. & Formula & $k_s^+$ (pred.) & $k_s^+$ (ref.) & Error [\%] \\
\hline
(1) & Na2008 & $k_s^+ = a^+\,[0.772 + ES\,(2.37 - ES)]$ & 80.4 & 176.1 & $-54.3$ \\
(2) &  &  & 195.1 & 361.9 & $-46.1$ \\[4pt]
(1) & Ch2015 & $k_s = 7.3\,S_a\,ES^{0.45}$ & 154.4 & 176.1 & $-12.3$ \\
(2) &  &  & 384.7 & 361.9 & $+6.3$ \\[4pt]
(1) & Sa2015 & $k_s = 2\,k$ & 243.4 & 176.1 & $+38.2$ \\
(2) &  &  & 501.4 & 361.9 & $+38.6$ \\[4pt]
(1) & Fo2017 & $k_s = k\,(0.67\,S_k^2 + 0.93\,S_k + 1.3)[1.07\,(1 - e^{-3.5\,ES})]$ & 101.3 & 176.1 & $-42.8$ \\
(2) &  &  & 262.1 & 361.9 & $-27.6$ \\[4pt]
(1) & FS2020 & $k_s = 2.11\,S_q$ & 90.8 & 176.1 & $-48.4$ \\
(2) &  &  & 186.6 & 361.9 & $-48.4$ \\
\end{tabular}
\end{ruledtabular}
\end{table}

These results confirm that the wavy geometries with $\overline R/k =
2, 3.3$ and $5$ operate within the fully rough regime. The consistency
with canonical scaling laws, including those of \citet{nikuradse1933} and
\citet{Colebrook1939}, reinforces the robustness of $\Delta U^+$ and
$k_s^+$ as general descriptors of rough-wall turbulence, even in
geometries that are large-scale, deterministic, and spatially
coherent. This trend also highlights the limitations of purely
geometric characterizations and supports the use
of velocity-based metrics as a more reliable, flow-informed measure of
equivalent roughness.

\subsubsection{\label{sec:meanvel_turb}Mean velocity}

\begin{figure}[t]
\includegraphics[width=\columnwidth]{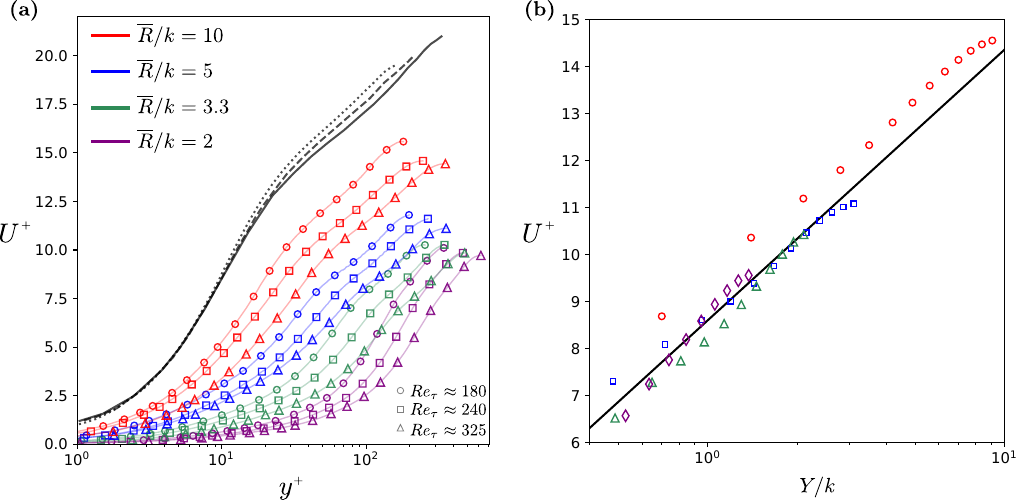}
\caption{(a) Mean streamwise velocity profiles ($U^+$) as a function of the
wall-normal coordinate ($y^+$). Rough-wall cases: $Re_\tau \approx
180$ ($\circ$), $240$ ($\square$), and $325$
($\triangle$). Smooth-pipe reference: $Re_\tau \approx 180$
($\cdots$), $240$ ($--$), and $325$ ($-$).
(b) Outer-scaled mean velocity profiles for the rough-wall cases at
$Re_\tau \approx 325$, plotted as a function of $Y/k_s$, where $Y$ is
the coordinate with respect to the wall-offset, see Appendix~\ref{app:offset}. The solid
black line shows the log-law reference for the roughest pipe
with $\overline R/k = 2$.}
\label{re_study_2}
\end{figure}

In Figure~\ref{re_study_2}(a), the mean streamwise velocity profiles
$U^+$ are shown for the different roughness levels and wall Reynolds
numbers. The profiles are plotted against the inner-scaled wall-normal
coordinate ${y}^+$. Increasing roughness (decreasing $\overline{R}/k$)
leads to a progressive downward shift of the profiles in inner units,
consistent across the range of Reynolds numbers considered. The
downward shift compared to the smooth pipe profile is quantified by
the roughness function discussed in the previous section.
A consistent feature observed in the mean velocity profiles is the
relatively weak sensitivity of $U^+$ to the Reynolds number within the range
investigated, particularly at higher roughness levels, which
indicates that turbulence dynamics are primarily governed
by the wall geometry once roughness effects become significant. This
behavior is consistent with the transition towards the fully rough
regime described by \citet{jimenez2004}, where the flow becomes
largely independent of viscous scaling and dominated by pressure drag
induced by the surface roughness. The systematic downward shift of the
profiles with increasing roughness intensity shown in panel (a)
further supports this interpretation as also
observed by \cite{saha_2015}. In their study, the mean velocity
profiles were obtained only up to the minimum wall radius
$R_{\min}$. This implies that the averaging excludes the crest regions
of the corrugation, effectively shifting the virtual origin of the
wall inward. In contrast, the average applied here, see
Eq.~\eqref{eq:average}, retains the full wall-normal
extent, resolving the velocity profile all the way to the local wall,
including trough regions, which allows for the characterization
of the roughness-induced deceleration in the viscous sublayer.

For a given $Re_\tau$, increasing wall
modulation reduces near-wall and core velocities due to
enhanced drag and momentum redistribution. These behaviors are
consistent with the roughness-controlled regime described by
\citet{Flack2014}, where surface geometry, not Reynolds number,
dominates the flow structure once the roughness scale exceeds a
critical threshold. In Figure~\ref{re_study_2}(b), the mean velocity
profiles are plotted as a function of the roughness-scaled wall-normal
coordinate $Y/k_s$ referred to the virtual origin, see Appendix~\ref{app:offset}. Only results corresponding
to the highest Reynolds number, $Re_\tau \approx 325$, are
shown. When scaled in this way, the profiles for $\overline R/k = 2,
3.3$ and $5$ collapse in the outer region, indicating
the self-similar behavior~\eqref{eq:Urough} with respect to $k_s$. The
solid black line represents the log-law reference for the case
$\overline R/k = 2$, for which we obtain the intercept $B = 8.6$.
The collapse of the profiles demonstrates that $k$, defined as
the peak-to-peak amplitude of the wall modulation, provides an
appropriate characteristic length scale for wavy-wall flows within
the present Reynolds-number range. The observed collapse further
suggests that the outer-layer dynamics are predominantly governed by
the geometric roughness rather than viscous effects, consistent with a
roughness-controlled turbulent regime.

\section{\label{sec:conclusions}Conclusions}

This study examined turbulent flow through pipes with sinusoidally
wavy walls using direct numerical simulations at moderate Reynolds
numbers. The analysis explored how wall amplitude modifies the mean
flow and drag, focusing on the friction factor and inner-scaled velocity profiles.
Waviness impacts quantitatively the flow behaviors in all flow
regimes, laminar, transitional and turbulent, as well as the
boundaries between these regimes.

In the laminar regime, the friction factor exceeds the
Hagen-Poiseuille prediction by 5--40\%. This excess disappears only
when an effective hydraulic diameter, extracted a posteriori from the
simulated pressure loss, is used to redefine the Reynolds number and
friction factor. The increased drag originates from a stable
recirculation bubble forming in the expanding half-wave at $Re_b
\approx 500$ for $\overline{R}/k \le 5$. As roughness increases, these
recirculations appear at decreasing Reynolds numbers, thickening the
near-wall region, displacing high-shear layers towards the core, and
promoting an early, geometry-induced transition to turbulence at
Reynolds numbers between 500 and 1000 typically assumed for
karst conduits \cite{worthington2017}. That is, the transitions occur
well below the threshold for smooth and rough pipes.

Between the laminar and fully rough turbulent regimes, the flow
exhibits a transitional state governed by finite-amplitude effects.
The onset of sustained turbulence is controlled by the balance between
the disturbance imposed by wall waviness and the increasing
susceptibility of the laminar flow to such perturbations as Reynolds
number increases. The observed transition boundary follows an
approximate power-law scaling between wall amplitude and Reynolds
number ($k/R_e\sim Re^{-\gamma}$ with $\gamma$ between $5/4$ and $3/2$),
consistent with a subcritical, bypass-type transition mechanism. This
behavior highlights the role of deterministic wall
geometry as an intrinsic source of finite disturbances, capable of
triggering and sustaining turbulence independently of external
forcing.

Once turbulence is established, the velocity profiles show little
sensitivity to Reynolds number and collapse according to roughness
level. This behavior is consistent with the fully rough regime, where
pressure drag dominates and the friction factor becomes almost
independent of Reynolds number \cite{jimenez2004, Flack2014}. The
reduction in both near-wall and core velocities with increasing
amplitude confirms the pressure-driven nature of the flow.
Purely geometric estimates of $k_s^+$ based on wall amplitude and
friction velocity tend to underpredict $\Delta U^+$ at high roughness
levels, though the discrepancy is modest. The small ratio $k^+_{s}/k^+
\approx 1.2$ indicates that the geometric peak-to-peak amplitude ($k$) remains the
best single descriptor of roughness in this configuration. The
remaining deviation reflects additional form drag, flow separation,
and axial pressure gradients, which depend on curvature rather than
amplitude alone. Nevertheless, the geometric amplitude captures the
main physical contribution to hydraulic roughness. The roughness effect
generated by the coherent wall shape is more intense than that of the
sand-grain or moderate amplitude random roughness \cite{nikuradse1933, demaio}.

Overall, the results confirm that $\Delta U^+$ and $k_s^+$ are
reliable descriptors of wall-induced drag once the flow is known. They
collapse data across Reynolds numbers and roughness levels, even in
structured, non-random geometries. The wavy pipe therefore provides a
clean and controllable test case for separating geometric and
stochastic effects. Its agreement with classical scaling laws supports
the use of deterministic sinusoidal geometries as simplified analogues
for complex rough surfaces. At the same time, our findings highlight
the limits of models for the friction factor based on a single roughness
parameter. As shown here, a friction factor for strong wall
modulations needs to reflect not only the sandgrain roughness that
determines the fully turbulent drag, but also the roughness dependent
critical Reynolds number for the laminar-turbulent transition as well as the
effective hydraulic radius that determines the laminar regime. 


In conclusion, classical characterizations of flow in rough conduits
based on the Moody diagram, or analytical correlations for the
Darcy-Weisbach friction factor need to be reassessed for conduits with
strong wall modulations \cite{flack2018moving}. This refers to the
definition of hydrodynamic concepts like the effective hydraulic diameter
in the laminar and the sandgrain roughness in the turbulent regimes as
well as the definition of the boundaries between these regimes, and
their relations to geometrical or predictable hydraulic properties.

\begin{acknowledgments}
The authors acknowledge funding from the European
Union (ERC, KARST, 101071836). The authors
report no conflict of interest.
\end{acknowledgments}

\appendix

\section{\label{app:offset}Wall offset and roughness function}

To enable a consistent comparison of the mean velocity in the core
region across roughness amplitudes, we introduce a wall offset or
virtual origin $\ell$. In rough-wall flows, the turbulent motions do
not perceive the geometric wall boundaries, but rather an effective
origin located somewhere within or above the roughness layer; its
definition is geometry-dependent and no universal formulation exists
in the literature \cite{raupach1991, saha_2015, chung2021}. For the
present wavy geometry, the effective hydraulic radius $R_e$ is the
natural hydrodynamic length scale, as it already governs the laminar
flow resistance (Section~\ref{sec:hydrad}), ensuring consistency
across flow regimes. We therefore define
\begin{equation}
    \ell = k \frac{R_e}{R_{\max}},
\end{equation}
where $R_e/R_{\max}$ weights the roughness amplitude $k$ by the
hydrodynamic effectiveness of the wall modulation. In the smooth-wall
limit, $k \to 0$ and $R_e \to R_{\max}$, so $\ell \to 0$ and the
standard wall-normal coordinate is recovered. The shifted coordinate
$Y = y - \ell$ aligns the outer-layer reference plane across
roughness amplitudes and allows for a consistent evaluation of the
roughness function $\Delta U^+$. The proposed offset yields a
reasonable collapse of the outer-scaled profiles for the moderate
and large roughness amplitudes ($\bar{R}/k = 2$, 3.3 and 5), as
shown in Figure~\ref{re_study_2}(b). The weakest roughness case ($\bar{R}/k = 10$)
exhibits a residual offset, which is expected given that its
roughness function $\Delta U^+$ differs significantly from the
strongly modulated cases, placing it in a different roughness regime.
This confirms that the proposed scaling captures the dominant
hydrodynamic effect of the wall modulation, without claiming a
perfect collapse across all amplitudes.

\section{\label{app}Second-order statistics}

\subsection{Transitional regime}
\begin{figure}[t]
\includegraphics[width=\columnwidth]{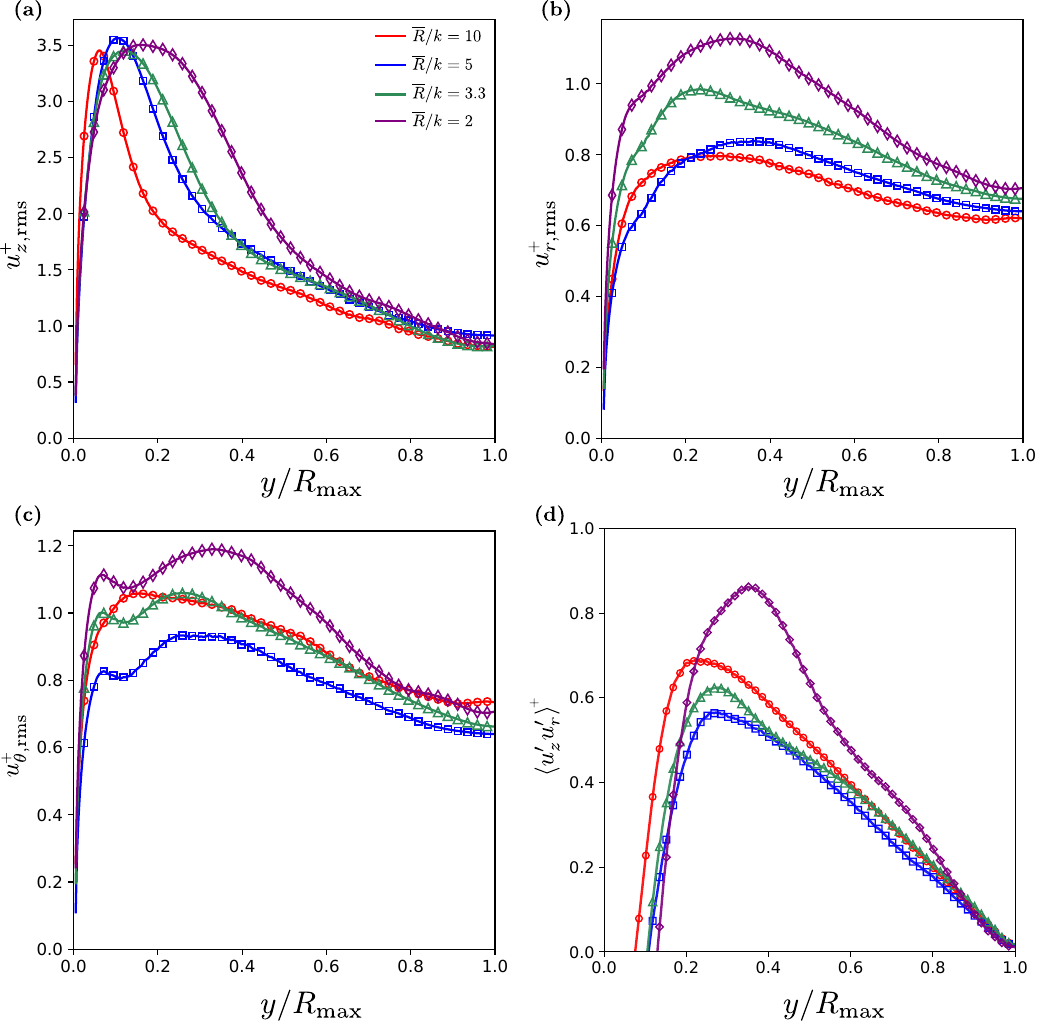}
\caption{Root-mean-square velocity fluctuation profiles: (a) axial ($u_z'^+$), (b) radial ($u_r'^+$), and (c) azimuthal ($u_{\theta}'^+$) velocities for different roughness heights at $Re_\tau \approx 120$. Panel (d) shows the corresponding Reynolds shear stress ($\langle u_r'u_z' \rangle^+$). Results are presented in outer scaling, with $y/R_{\max}$ as the wall-normal coordinate.}
\label{ks_2000}
\end{figure}

Figure~\ref{ks_2000} shows the RMS velocity fluctuations in the axial
($u_z'^+$), radial ($u_r'^+$), and azimuthal ($u_\theta'^+$)
directions, together with the Reynolds shear stress ($\langle u_r'u_z'
\rangle^+$) for different roughness heights at $Re_\tau \approx 120$,
expressed in outer units ($y/R_{\max}$). The axial RMS peaks near the
wall for all cases, with little variation outside the wavy sublayer.
Radial and azimuthal fluctuations increase with roughness, with the
radial component showing a mild outward shift of the peak and the
azimuthal component developing a broader outer shoulder. The Reynolds
shear stress profiles collapse in the outer region, indicating that
large-scale momentum transfer remains similar to the smooth-wall case
despite near-wall geometric modulation. Near the wall, however, the
Reynolds shear stress exhibits a finite value at $y/R_{\max} \approx
0$ and becomes slightly negative before reaching zero at the physical
wall. This behaviour reflects the presence of recirculation zones at
this transitional Reynolds number, where locally reversed flow
contributes negative correlations between the radial and axial
velocity fluctuations. At higher $Re_\tau$ (Figure~\ref{re_study_variance}), the flow is 
fully turbulent and the turbulent signal dominates, rendering this 
near-wall feature negligible in the global average.

\subsection{Turbulent regime}
\begin{figure}[b]
\includegraphics[width=\columnwidth]{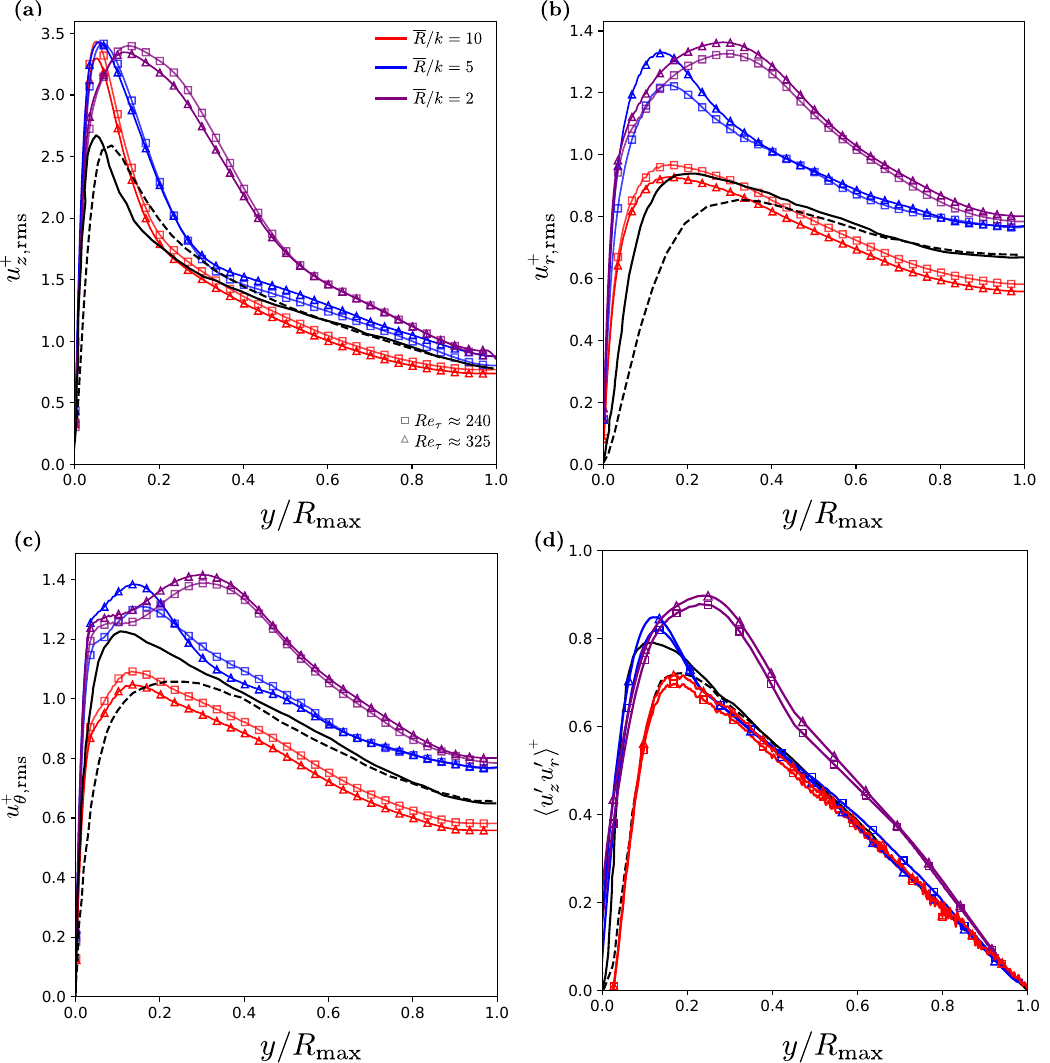}
\caption{Root-mean-square velocity fluctuation profiles: (a) axial, (b) radial, and (c) azimuthal components for different roughness heights. Panel (d) shows the corresponding Reynolds shear stress ($\langle u_r' u_z' \rangle^+$). Results are shown for Reynolds numbers $Re_\tau \approx 240$ ($\square$) and $325$ ($\triangle$), across the investigated roughness levels. The black lines denote the smooth-wall pipe case, with $Re_\tau \approx240$ ($--$), and $325$ ($-$).}
\label{re_study_variance}
\end{figure}

Figure~\ref{re_study_variance} presents the RMS and Reynolds stress
profiles at higher Reynolds numbers, $Re_\tau \approx 240$ and
$325$. The axial fluctuations display near-wall peaks that grow and
slightly shift outward with increasing roughness, while the radial and
azimuthal components show monotonic amplification near the wall and
moderate outward extension. The Reynolds shear stress deviates from
the linear total-stress distribution close to the wall, where viscous
contributions are important, but approaches the expected smooth-wall
linear trend further from the wall. For the largest wall amplitude of
$\overline R/k = 2$, the linear trend can be seen only in the central
core region. Overall, the results indicate that increasing roughness 
enhances near-wall turbulence intensities in all velocity components, 
with the strongest effects observed in the streamwise and radial 
fluctuations. The weakest roughness case ($\overline{R}/k = 10$), 
which falls in the transitionally rough regime, shows near-wall 
fluctuations comparable to or slightly below the smooth-wall 
reference, consistent with the limited turbulence enhancement 
expected when roughness effects remain partially damped by viscosity.
Notably, the near-wall profiles of the wavy cases show little 
sensitivity to $Re_\tau$, with the statistics at $Re_\tau \approx 240$ 
and $325$ remaining close to each other for each roughness level. 
This is consistent with fully rough flow, in which the viscous 
sublayer is destroyed by the wall modulation and near-wall turbulence 
is governed by the geometry rather than viscosity. By contrast, the 
smooth-wall profiles exhibit a clear $Re_\tau$ dependence, with the 
near-wall peak growing and shifting with increasing Reynolds number, 
as expected from classical viscous scaling. While the amplitude and 
spatial distribution of velocity fluctuations are significantly 
affected, the Reynolds shear stress remains comparatively robust when 
expressed in wall units, highlighting the dominant role of wall shear 
in scaling turbulent momentum transfer.

%

\end{document}